\documentclass[12pt]{article}

\input{epsf}
\usepackage{cite}
\usepackage[dvips]{graphicx}
\usepackage{amsmath}
\usepackage{epsfig}
\usepackage{amsfonts}
\usepackage{amssymb}
\usepackage{multirow}
\usepackage{array}

\usepackage{axodraw4j}
\usepackage{pstricks}
\usepackage{color}

\usepackage{epsfig}
\usepackage{amsmath}
\usepackage{amssymb}

\usepackage{subfigure}

\newcommand{\be}{\begin{equation}}
\newcommand{\ee}{\end{equation}}
\newcommand{\bee}{\begin{equation*}}
\newcommand{\eee}{\end{equation*}}
\newcommand{\bea}{\begin{eqnarray}}
\newcommand{\eea}{\end{eqnarray}}
\newcommand{\bean}{\begin{eqnarray*}}
\newcommand{\eean}{\end{eqnarray*}}

\usepackage{amsfonts}
\usepackage{amssymb}
\usepackage{amsmath}
\usepackage{mathrsfs}
\usepackage{latexsym}
\usepackage{cancel}
\usepackage{bm}
\usepackage{bbm}
\usepackage[mathcal]{euscript}
\usepackage{graphicx}
\usepackage{color}

\begin{document}

\setcounter{page}{0}
\thispagestyle{empty}

\begin{flushright}
CERN-PH-TH/2010-187\\
CP3-10-34\\
\today
\end{flushright}

\vskip 8pt

\begin{center}
{\bf \LARGE {
Non-resonant New Physics \\
\vskip 8pt
in Top Pair Production at Hadron Colliders \\
 }}
\end{center}

\vskip 12pt

\begin{center}
{\bf C\'eline Degrande$^{a}$, Jean-Marc G\'erard$^{a}$, Christophe Grojean$^{b,c}$,}\\
 \vskip 6pt
{\bf    Fabio Maltoni$^{a}$, }
{\bf and G\'eraldine  Servant $^{b,c}$ }
\end{center}

\vskip 20pt

\begin{center}

\centerline{$^{a}${\it
Centre for Cosmology, Particle Physics and Phenomenology (CP3),}}
\centerline{\it Universit\'e catholique de Louvain, B-1348, Louvain-la-Neuve, Belgium}
\centerline{$^{b}${\it CERN Physics Department, Theory Division, CH-1211 
Geneva 23, Switzerland}}
\centerline{$^{c}${\it Institut de Physique Th\'eorique, CEA/Saclay, F-91191 
Gif-sur-Yvette C\'edex, France}}
\vskip .3cm
\centerline{\tt  celine.degrande@uclouvain.be,jean-marc.gerard@uclouvain.be}
\centerline{\tt christophe.grojean@cern.ch,fabio.maltoni@uclouvain.be,geraldine.servant@cern.ch}
\end{center}

\vskip 13pt

\begin{abstract}
\vskip 3pt
\noindent
We use top quark pair production as a probe of {\it top-philic} non-resonant new physics. Following a low energy effective field theory approach, we calculate several key observables in top quark pair production at hadron colliders ({\it e.g.}, total cross~section, $t\bar{t}$ invariant mass distribution, forward-backward asymmetry, spin correlations) including the interference of the Standard Model with dimension-six operators. We determine the LHC reach in probing new physics after having taken into account the Tevatron constraints. In particular, we show that the gluon fusion process $gg\rightarrow t\bar{t}$ which remains largely unconstrained at the Tevatron is affected by only one {\it top-philic} dimension-six operator, the chromo-magnetic moment of the top quark.
This operator can be further constrained by the LHC data as soon as a precision of about 20\% is reached 
for the total $t\bar{t}$ cross-section. 
While our approach is general and model-independent, it is particularly relevant to models of Higgs and top compositeness, which we consider in detail, also in connection with $t\bar{t}t\bar{t}$ and $t\bar{t}b\bar{b}$ production. 
\end{abstract}

\newpage

\tableofcontents

\vskip 13pt

\section{Introduction}

Top quark physics is among the central physics topics at the Tevatron and it will remain so at the Large Hadron Collider (LHC) in the next few years. The top being the only quark with a coupling to the Higgs of order one, it is expected to play a special role in electroweak symmetry breaking and as a result its coupling to new physics could be large. Searching for beyond the Standard Model (SM) physics in observables involving the top quark is therefore strongly motivated.
Top physics has actually reached a high level of sophistication and we already know a lot from the Tevatron which has  set strong constraints on {\it top-philic}  new physics~\cite{Beneke:2000hk,Bernreuther:2008ju, Han:2008xb, Frederix:2010cn}. Until recently, Tevatron was the only source of top quarks. However, LHC has finally produced its first top quarks~\cite{cms:ichep, atlas:ichep, cms:hcp, atlas:hcp} and  will soon become a major top quark factory. 

So, what more can we expect to learn about the top at the LHC? The top quark is mainly produced in $t \overline{t}$ pairs, the single top rate being roughly one third. At the lowest order in perturbative QCD, this occurs either from quark-antiquark annihilation, $q\overline{q} \to t\bar{t}$,  or from gluon fusion, $gg\to t\bar{t}$. What has been mainly tested at the Tevatron is the hard process $q\overline{q}\rightarrow t\overline{t}$, which contributes to 85$\%$ of the total  $t\bar{t}$ cross-section. The new physics contributions to $gg \rightarrow t \overline{t}$  remain to date largely unconstrained. Therefore,  there is a large  unexplored territory related  the top pair production that we can hope to unveil at the LHC, where $t\bar{t}$ is produced at 90$\%$ (70$\%$) by gluon fusion at 14 TeV (7 TeV). This simple observation is the starting point of our investigation.

A large effort  has been devoted to search for new physics in  $t \overline{t}$  resonances~\cite{Barger:2006hm, Choudhury:2007ux, Frederix:2007gi}.  While the current existing bounds do not forbid the existence of new degrees of freedom that are within the kinematical reach of the Tevatron and the LHC, electroweak precision data~\cite{:2005ema} together with constraints from flavour physics make plausible if not likely that there exists a mass gap between the SM degrees of freedom and any new physics threshold. In this case, the effects of new physics on a SM process like the $t\bar{t}$ production can be well captured by higher dimensional interactions among the SM particles. These new interactions are assumed to respect all the symmetries of the SM. In this work, we follow this low-energy effective field theory approach. Our study concentrates on testing non-resonant {\it top-philic} new physics, focusing on modifications from new physics to top pair production. The study of some dimension-six operators on $t\bar{t}$ production at the Tevatron was initiated in Refs. \cite{Hill:1993hs, Atwood:1994vm,Cheung:1995nt,Whisnant:1997qu,Hikasa:1998wx} 
and further explored in Refs~\cite{Lillie:2007hd,Kumar:2009vs,Jung:2009pi,Hioki:2009hm,Choudhury:2009wd}. An effective field theory approach to single top production would be the subject of another study and we refer the reader to Ref.~\cite{Zhang:2010px,Zhang:2010dr,AguilarSaavedra:2010zi} for this purpose. On the other side, the effect of higher dimensional operators on top anomalous couplings has already been discussed in Refs.~\cite{Grzadkowski:2003tf,Grzadkowski:2008mf,
AguilarSaavedra:2008zc, AguilarSaavedra:2009mx, AguilarSaavedra:2010nx}.
 
This work is organized as follows. We derive the effective Lagrangian relevant for top quark pair production, considering operators up to dimension-six, and connect it with composite and resonant models in Section~\ref{sec:EffLag}. Section~\ref{sec:ttbar} is devoted to top pair production. Our goal is to present an analysis where the full parameter space is considered with respect to the  constraints from the Tevatron. Once we have identified the unconstrained region, we determine which region will be further probed at the LHC. In particular, we include the latest CDF measurement of the invariant mass differential cross-section (based on 4.8~$\textrm{fb}^{-1}$ of data~\cite{Goldschmidt:2010}) in Section~\ref{sec:distributions}. We explore different observables like the $t\bar{t}$ invariant mass, the $p_T$ and the rapidity distributions at the LHC and find that their discriminating power is modest. Our final results are shown in Figs.~\ref{Mttconstraints} and~\ref{SummaryPlot}. We also show that  the forward-backward asymmetry at the Tevatron is affected by only one particular linear combination of four-fermion operators, that remains unconstrained by cross section measurements (Section~\ref{sec:AFB}).  Finally,  spin correlations, evaluated in Section~\ref{sec:spins}, are found to be quite sensitive to new physics. Another complementary and compelling probe of {\it top-philic} new physics is the four-top production~\cite{Pomarol:2008bh}. This is  discussed in Section~\ref{sec:4t} together with top pair production in association with $b\bar{b}$. We summarize our results and conclude in Section~\ref{sec:conclusion}.

\section{Effective Lagrangian for top quark pair production}
\label{sec:EffLag}

\subsection{Dimension-six operators with two top quarks}\label{sec:dim6}

When working with an effective field theory, the starting point is to consider the underlying symmetries.  Here, we assume that the symmetries of the SM, including the baryon number conservation, are unbroken by the new physics. The gauge invariant operators of dimension-six built from the SM degrees of freedom were classified many years ago in Ref.~\cite{Buchmuller:1985jz} and reconsidered recently in Ref.~\cite{Grzadkowski:2010es}. We shall focus our analysis on {\it top-philic} new physics, \textit{i.e.}, new physics that manifests itself in the top sector, as well-motivated in a large class of theories to be discussed in Section~\ref{sec:Composite}, but we do not consider new interactions that would only affect the standard gluon vertices like for instance  the interactions generated by the operator $\mathcal{O}_G=f_{ABC} G^A_{\mu\nu} G^{B\, \nu\rho} G^C_{\rho}{}^\mu$ (see Refs.~\cite{Cho:1993eu, Cho:1994yu, Simmons:1995hb, Zhang:2010dr} for a study of its effects on top pair production).  Hence we consider the set of operators which affect the $t\bar{t}$ production at tree-level by interference with the SM amplitudes. Both at the Tevatron and at the LHC, the dominant SM amplitudes are those involving QCD in quark-antiquark annihilation or gluon fusion. Therefore we shall neglect all new interactions that could interfere only with SM weak processes like $q\bar{q}\to Z(\gamma)\to t\bar{t}$. Our analysis aims at identifying the effects of the new physics on top pair production, so it ignores the operators which affect the decay of the top~\cite{Zhang:2010dr, AguilarSaavedra:2008zc, Drobnak:2010ej}. We are then left with only two classes of dimension-six gauge-invariant operators~\cite{Buchmuller:1985jz}: 

\begin{itemize}
 \item operators with a top and an antitop and one or two gluons, namely
\begin{eqnarray}
 \mathcal{O}_{gt} &=& \left[ \bar{t} \gamma^\mu T^A D^\nu t \right] G^A_{\mu\nu}\, , \nonumber\\
 \mathcal{O}_{gQ} &=& \left[ \bar{Q} \gamma^\mu T^A D^\nu Q \right] G^A_{\mu\nu}\, , \nonumber\\
 \mathcal{O}_{hg} &=& \left[ \left( H \bar{Q} \right) \sigma^{\mu\nu} T^A t \right] G^A_{\mu\nu}\, ,  \label{eq:gtt}
\end{eqnarray}
where $Q=(t_L,b_L)$ denotes the left-handed weak doublet of the third quark generation, $t$ is the right-handed top quark, $T^A$ are the generators of $SU(3)$ in the fundamental representations normalized to $\textrm{tr} (T^A T^B) = \delta^{AB}/2$.

 \item four-fermion operators with a top and an antitop together with a pair of light  quark and antiquark that can be organized following their chiral structures:\vspace{3mm}\\
$\bar{L}L\bar{L}L$:
 \begin{eqnarray}
 \mathcal{O}_{Qq}^{(8,1)} &=& \!\left( \bar{Q} \gamma^\mu T^A Q \right) \!\left( \bar{q} \gamma_\mu T^A q \right), \nonumber\\
\mathcal{O}_{Qq}^{(8,3)} &=& \!\left( \bar{Q} \gamma^\mu T^A \sigma^I Q \right) \!\left( \bar{q} \gamma_\mu T^A \sigma^I q \right),
\label{eq:LLLL}
 \end{eqnarray}
$\bar{R}R\bar{R}R$:
 \begin{eqnarray}
  \mathcal{O}_{tu}^{(8)} &=& \!\left( \bar{t} \gamma^\mu T^A t \right) \!\left( \bar{u} \gamma_\mu T^A u \right) , \nonumber\\ 
  \mathcal{O}_{td}^{(8)} &=& \!\left( \bar{t} \gamma^\mu T^A t \right) \!\left( \bar{d} \gamma_\mu T^A d \right) ,
\label{eq:RRRR}
 \end{eqnarray}
$\bar{L}L\bar{R}R$:
 \begin{eqnarray}
  \mathcal{O}_{Qu}^{(8)} &=& \!\left( \bar{Q} \gamma^\mu T^A Q \right) \!\left( \bar{u} \gamma_\mu T^A u \right) ,\nonumber\\
  \mathcal{O}_{Qd}^{(8)} &=& \!\left( \bar{Q} \gamma^\mu T^A Q \right) \!\left( \bar{d} \gamma_\mu T^A d \right),\nonumber\\
  \mathcal{O}_{tq}^{(8)} &=& \!\left( \bar{q} \gamma^\mu T^A q \right) \!\left( \bar{t} \gamma_\mu T^A t \right),
\label{eq:LLRR}
 \end{eqnarray}
$\bar{L}R\bar{L}R$:
 \begin{eqnarray}
  \mathcal{O}_{d}^{(8)} &=& \!\left( \bar{Q} T^A t \right) \!\left( \bar{q} T^A d \right),
 \label{eq:LRLR}
\end{eqnarray}
where $\sigma^I$ are the Pauli matrices (normalized to $\textrm{tr} (\sigma^I \sigma^J) = 2 \delta^{IJ}$), $q$ and $u$ and $d$ are respectively the left- and right-handed components of the first two generations. 
\end{itemize}

Note that there also exist some colour-singlet analogues of all these operators, that can be generated by a heavy $Z'$ in $s$-channel for example. However, they do not interfere with the SM QCD amplitudes and therefore are not considered here. In other words, we assume that new physics manifests itself at low energy only through operators interfering with the SM. We will comment further on this assumption in Section \ref{sec:validity}.
All the four-fermion operators are written in the mass-eigenstates basis and no CKM mixing will enter in our analysis since we are neglecting weak corrections. We have discarded the operator $\left(\bar{Q} T^A \gamma^\mu q_L\right) \left(\bar{Q} T^A \gamma^\mu q_L\right)$ as well as its $SU(2)$ triplet and color singlet\footnote{Here, the color singlet should also be considered in principle since the color flow is in the t-channel.} analogues since they are already strongly constrained by flavour physics~\cite{Bona:2007vi}. Note also that operators with a different Lorentz or gauge structure, like for instance $(\bar{Q} \gamma^\mu T^A q) (\bar{q} \gamma^\mu T^A Q)$ or $(\bar{t} \gamma^\mu T^A u) (\bar{u} \gamma_\mu T^A t)$, can be transformed (using Fierz identities, see App.~\ref{app:Fierz}) into linear combinations of the four-fermion operators listed above  and their colour-singlet partners. 

The $\bar{L}R\bar{L}R$ operator $\mathcal{O}_d^{(8)}$ involves both the left- and the right-handed components of the down quark. Given the fact that QCD interactions are chirality-diagonal, it can only interfere with the SM amplitude after a mass insertion. Therefore, its contribution to the $t\bar{t}$ production cross section is negligible and we shall not consider it further in our analysis.

It is rather natural to assume the universality of new physics with respect to the light generations. In that limit, the contribution to the cross~section from the second generation is more than two orders of magnitude smaller than the one from the first generation due to the different parton distribution functions (pdf). We shall therefore concentrate on the contribution from the first generation only.

Our list (\ref{eq:gtt})--(\ref{eq:LRLR}) of top-philic operators contains eleven operators. However, they are still not all independent. Indeed, using the equation of motion for the gluons:
\begin{equation}
D^\nu G_{\mu\nu}^A = g_s\sum_f \bar{f}\gamma_\mu T^A f,\label{motion}
\end{equation}
we obtain the following two relations :
\begin{eqnarray}
&&
\mathcal{O}_{gt} + \mathcal{O}_{gt}^\dagger = -g_s \sum_{{\textrm{ generations}}} 
\left( \mathcal{O}_{tq}^{(8)} + \mathcal{O}_{tu}^{(8)}  + \mathcal{O}_{td}^{(8)} \right) , \\
&&
\mathcal{O}_{gQ} + \mathcal{O}_{gQ}^\dagger = -g_s \sum_{{\textrm{ generations}}} 
\left( \mathcal{O}_{Qq}^{(8,1)} + \mathcal{O}_{Qu}^{(8)}  + \mathcal{O}_{Qd}^{(8)} \right).
\end{eqnarray}
The linear combinations $\mathcal{O}_{gt} - \mathcal{O}_{gt}^\dagger$ and $\mathcal{O}_{gQ} - \mathcal{O}_{gQ}^\dagger$ do not interfere with the SM amplitudes because the associated vertices are CP-odd and we are not concerned about CP violating observables (see Ref.~\cite{Zhang:2010dr,Grzadkowski:1997yi} for a discussion on possible observables sensitive to CP violation). Consequently, the two operators $\mathcal{O}_{gt}$ and $\mathcal{O}_{gQ}$ can be dropped in our analysis and only one two-fermion operator, namely $\mathcal{O}_{hg}$, interferes with the SM gluon fusion process!

In conclusion, the most general top-philic Lagrangian that can affect the $t\bar{t}$ production involves eight dimension-six operators
\begin{equation}
		\label{eq:ttlag1}
{\mathcal L}_{t\bar{t}}\left(\Lambda^{-2}\right)  =  
 \frac{1}{\Lambda^2} \left( 
\left(c_{hg} {\mathcal O}_{hg} + h.c. \right)
+ \sum_i c_i {\mathcal O}_i
\right),
\end{equation}
where $i$ runs over the seven self-hermitian four-fermion operators of Eqs.~(\ref{eq:LLLL})--(\ref{eq:LLRR}).

In Eq.~\eqref{eq:ttlag1}, the coefficient $c_{hg}$ might be  complex. However, since we are concerned with CP-invariant observables, only its real part enters in the interference with the SM processes and therefore we shall assume in our analysis that $c_{hg}$ is real. This coefficient corresponds to a chromomagnetic moment for the top quark.

\subsection{The relevant operators}
\label{sec:relop}

In Eq.~\eqref{eq:ttlag1}, we have identified eight independent top-philic operators. 
Yet, additional simple considerations are going to show that physical observables like the $t\bar{t}$ production total cross section, the $m_{t\bar{t}}$ invariant-mass distribution or the forward-backward asymmetry, depend only on specific linear combinations of these operators.

The seven four-fermion operators can be combined to form linear combinations with definite $SU(2)$ isospin quantum numbers. In the isospin-0 sector, it is further convenient to define axial and vector combinations of the light quarks:
\begin{equation}
\mathcal{O}_{Rv} =  \mathcal{O}_{tu}^{(8)}+\mathcal{O}_{td}^{(8)}+ \mathcal{O}_{tq}^{(8)}, \\
\qquad
\mathcal{O}_{Ra} = \mathcal{O}_{tu}^{(8)}+\mathcal{O}_{td}^{(8)}-\mathcal{O}_{tq}^{(8)},
\end{equation}
and similar operators involving the left-handed top quarks:
\begin{equation}
\mathcal{O}_{Lv} =  \mathcal{O}_{Qu}^{(8)}+\mathcal{O}_{Qd}^{(8)}+ \mathcal{O}_{Qq}^{(8,1)}, \\
\qquad
\mathcal{O}_{La} = \mathcal{O}_{Qu}^{(8)}+\mathcal{O}_{Qd}^{(8)}-\mathcal{O}_{Qq}^{(8,1)}.
\end{equation}
The reason is that the axial operators  are asymmetric under the exchange of the quark and antiquark while the vector operators are symmetric\footnote{The matrices $C\gamma^\mu\gamma^5$ are antisymmetric but the matrices $C\gamma^\mu$ are symmetric, $C$ being the charge conjugation matrix.}:
\begin{equation}
\begin{array}{c}
 \left[ \bar{\psi}\left(k_1\right) \gamma^{\mu} \gamma^5T^A \psi \left(k_2\right) \right]
 =  -\left[ \bar{\psi}^c\left(k_2\right) \gamma^{\mu} \gamma^5T^A \psi^c\left(k_1\right) \right],
\\[.3cm]
\left[ \bar{\psi}\left(k_1\right) \gamma^{\mu} T^A \psi \left(k_2\right) \right]
 =  \left[ \bar{\psi}^c\left(k_2\right) \gamma^{\mu} T^A \psi^c\left(k_1\right) \right].
\end{array}
\end{equation}
Therefore, the interferences of $\mathcal{O}_{Ra}$ and $\mathcal{O}_{La}$ with the SM will be odd under the exchange of the momenta of the initial partons and these axial operators can only contribute to observables that are odd functions of the scattering angle and certainly not to the total cross section. On the contrary, the operators $\mathcal{O}_{Rv}$ and $\mathcal{O}_{Lv}$ are even functions of the scattering angle and can contribute to $\sigma_{t\bar{t}}$.

In addition, the operators ${\mathcal O}_{Rv}$ and ${\mathcal O}_{Lv}$ will obviously produce the same amount of top pairs but with opposite chirality. Consequently, the spin-independent observables associated to the $t\bar{t}$ production are expected to depend only on the sum $\mathcal{O}_{Rv}+\mathcal{O}_{Lv}$ while
the difference $\mathcal{O}_{Rv}-\mathcal{O}_{Lv}$ will only contribute to spin-dependent observables. Similarly, but with a sign flip, only their difference, $\mathcal{O}_{Ra}-\mathcal{O}_{La}$,
can contribute to spin-independent observables and in particular to the $t\bar{t}$ differential cross~section after summing over the spins.
The orthogonal combination $\mathcal{O}_{Ra}+\mathcal{O}_{La}$ could contribute to 
spin-dependent observables which are odd functions of the scattering angle, but we shall not consider any observable of this type in our analysis.

Therefore, we expect a dependence of the total pair production cross section on the sum
\begin{equation}
c_{Vv}=c_{Rv}+c_{Lv} \ \ \textrm{with} \ \ 
\left\{
\begin{array}{l}
c_{Rv}=c_{tq}/2+(c_{tu}+c_{td})/4\\
c_{Lv}=c^{(8,1)}_{Qq}/2+(c_{Qu}+c_{Qd})/4\\
\end{array}
\right.
\end{equation}
and the forward-backward asymmetry will depend on the combination
\begin{equation}
c_{Aa}=c_{Ra}-c_{La}  \ \ \textrm{with} \ \ 
\left\{
\begin{array}{l}
c_{Ra}=-c_{tq}/2+(c_{tu}+c_{td})/4\\
c_{La}=-c^{(8,1)}_{Qq}/2+(c_{Qu}+c_{Qd})/4.
\end{array}
\right. 
\end{equation}
The difference 
\begin{equation}
c_{Av}=c_{Rv}-c_{Lv}
\end{equation} 
can only contribute to spin-dependent observables (see Section~\ref{sec:spins}).

The isospin-1 sector is spanned by the three combinations:
\begin{equation}
\mathcal{O}_{Rr} =\mathcal{O}_{tu}^{(8)}-\mathcal{O}_{td}^{(8)}\, ,\
\mathcal{O}_{Lr} =\mathcal{O}_{Qu}^{(8)}-\mathcal{O}_{Qd}^{(8)}\  \textrm{ and } \
{\mathcal O}_{Qq}^{(8,3)}\, .
\end{equation}
Again, parity arguments lead to the conclusion that the total cross section can only depend on the combination 
\begin{equation}
c_{Vv}^\prime=(c_{tu}-c_{td})/2+(c_{Qu}-c_{Qd})/2+c_{Qq}^{(8,3)},
\label{eq:isospin1}
\end{equation}
while the forward-backward asymmetry will only receive a contribution proportional to
\begin{equation}
c_{Aa}^\prime=(c_{tu}-c_{td})/2-(c_{Qu}-c_{Qd})/2+c_{Qq}^{(8,3)}.
\end{equation}
and spin-dependent observables will depend on (see App.~\ref{app:xsec})
\begin{equation}
c_{Av}^\prime=(c_{tu}-c_{td})/2-(c_{Qu}-c_{Qd})/2-c_{Qq}^{(8,3)}.
\end{equation}

Numerically, we shall see in Section~\ref{sec:totalcs} that the isospin-0 sector gives a larger contribution to the observables we are considering than the isospin-1 sector. This is due to the fact that a sizeable contribution to these observables is coming from a phase-space region near threshold where the up- and down-quark contributions are of the same order.

It is interesting to note that, in composite models, where the strong sector is usually invariant under the weak-custodial symmetry $SO(4)\to SO(3)$~\cite{Agashe:2006at}, the right-handed up and down quarks certainly transform as a doublet of the $SU(2)_R$ symmetry, and therefore $c_{Qu}=c_{Qd}$. There are however various ways to embed the right-handed top quarks into a $SO(4)$ representation~\cite{Pomarol:2008bh}: if it is a singlet, then $c_{tu}=c_{td}$ also and the isospin-1 sector reduces to the operator $\mathcal{O}^{(8,3)}_{Qq}$ only.

\begin{figure}[b]
\centering
\subfigure[Chromomagnetic operator $\mathcal{O}_{hg}=(H\bar{Q})\sigma^{\mu \nu}  T^A t \ G^A_{\mu\nu}$]{
\includegraphics[scale=0.65]{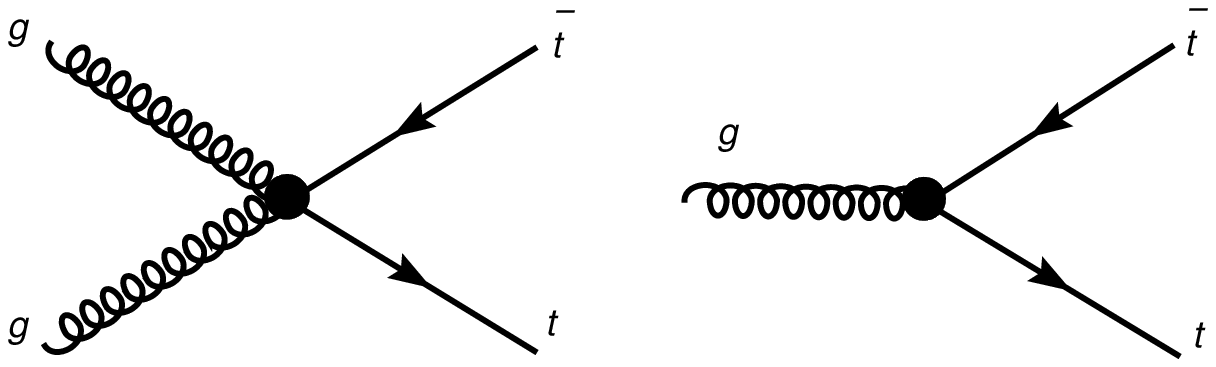}}
\hspace{2cm}
\subfigure[Four-fermion operators]{
\includegraphics[scale=.64]{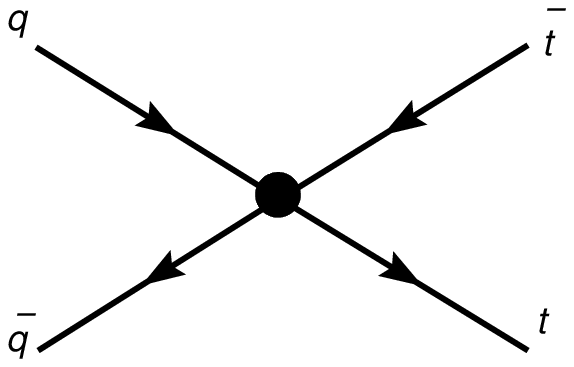}}
\caption{\label{fig:summaryoperators}
\small A Feynman representation of the  relevant operators  for $t\bar{t}$ production at hadron colliders.}
\end{figure}

In summary, the relevant effective Lagrangian for $t\bar{t}$ production contains a single two-fermion operator and seven four-fermion operators conveniently written as:
\begin{equation}
		\label{eq:ttlag}
{\mathcal L_{t\bar{t}}}  =  
+  \frac{1}{\Lambda^2} \!\left( 
\left(c_{hg} {\mathcal O}_{hg} + h.c.\right)
+ 
\left(c_{R\, v} {\mathcal O}_{R\, v} + c_{R\, a} {\mathcal O}_{R\, a} + c^{\prime}_{Rr} \mathcal{O}_{Rr}^\prime
+ R\leftrightarrow L\right) + c_{Qq}^{(8,3)} {\mathcal O}_{Qq}^{(8,3)}
\!\right).
\end{equation}

The vertices arising from the dimension-six operators given in Eq.~\eqref{eq:ttlag} relevant for top pair production at hadron colliders are depicted in Fig.~\ref{fig:summaryoperators}.

\subsection{Connection with composite top and resonance models}
\label{sec:Composite}

The effects of a composite top have been first studied in Ref.~\cite{Georgi:1994ha}. The construction of an effective Lagrangian for the fermionic sector has been discussed in details in Ref.~\cite{Pomarol:2008bh}. It relies on the assumption of partial compositeness, meaning that SM fermions are assumed to be linearly coupled to the resonances of the strong sector through mass mixing terms (see e.g \cite{Contino:2006nn}). The composite models are characterized by a new strong interaction responsible for the breaking of the electroweak symmetry and broadly parametrized by two parameters~\cite{Giudice:2007fh}: a dimensionless coupling $g_\rho$ and a mass scale $m_\rho$. The latter, associated with the heavy physical states, was generically denoted $\Lambda$ in Eq.~\eqref{eq:ttlag1}. In order to alleviate the tension with EW precision data, we assume that  in the limit where all the gauge and Yukawa interactions of the SM are switched off,  the full Higgs doublet is an exact Goldstone boson living in the $G/H$ coset space of a spontaneously broken symmetry of the strong sector. In such a case, $f$, the decay constant of the Goldstones, is related to $g_\rho$ and $m_\rho$ by
\begin{equation}
m_\rho = g_\rho f\, 
\end{equation}
with $1\lesssim g_\rho\lesssim4\pi$. The effective Lagrangian of the gauge and Higgs sectors was constructed in Ref.~\cite{Giudice:2007fh}. 	

At energies below the resonances masses, the dynamics of the top sector is described by the usual SM Lagrangian supplemented by a few higher dimensional operators. Simple rules control the size of these different operators, referred as Naive Dimensional Analysis (NDA)~\cite{Georgi:1992dw, Manohar:1983md}. Inspired by the rather successful chiral perturbation approach to QCD at low scale, NDA provides the following rules for the effective operators beyond ${\mathcal L}_{SM}$ in equation~\eqref{eq:ttlag1}:
\begin{enumerate}
	\item first, multiply by an overall factor $f^2$;
	\item then, multiply by a factor $\frac{1}{f}$ for each strongly interacting field;
	\item finally, multiply by powers of $m_\rho$ (instead of $\Lambda$) to get the right dimension.
\end{enumerate}

Hereafter, we consider two classes of gauge-invariant operators which are relevant  for top pair production: 
\begin{itemize}
\item                                             
Operators that contain only fields from the strong sector are called dominant because their coefficients scale like $g_\rho^2$. In most composite top models, only its right component is composite to avoid experimental constraints. In this case, there is only one such operator since the color octet equivalent is related to the color singlet by a Fierz transformation ($\mathcal{O}^{(8)}_{R}={\textrm{\small 1/3}}\, \mathcal{O}_{R}$),
\be
 \mathcal{O}_{R} = \!\left( \bar{t} \gamma^\mu t \right) \!\left( \bar{t} \gamma_\mu t \right).\label{odr}
\ee
If only the left handed top is composite, there are two independent dominant operators, 
\be
 \mathcal{O}_{L}^{(1)} = \!\left( \bar{Q} \gamma^\mu Q \right) \!\left( \bar{Q} \gamma_\mu Q \right) ,\qquad
 \mathcal{O}_{L}^{(8)} = \!\left( \bar{Q} \gamma^\mu T^A Q \right) \!\left( \bar{Q} \gamma_\mu T^A Q \right).\label{odl}
\ee
In the most general scenario where both chiralities are composite, two additional operators should also be considered,
\be
 \mathcal{O}_{S}^{(1)} = \!\left( \bar{Q} t \right) \!\left( \bar{t} Q \right), \qquad
 \mathcal{O}_{S}^{(8)} = \!\left( \bar{Q} T^A t \right) \!\left( \bar{t} T^A Q \right).\label{odb}
\ee
Needless to say that none of these operators contribute at tree-level to $t\bar{t}$ production. Yet they are relevant for direct production of four top-quarks (see Section~\ref{sec:4t}).
\item 
Operators which contribute directly to $t\bar{t}$ production are subdominant. On the one hand, the four-fermion operators given in Eq.~\eqref{eq:ttlag} contain at most two fields from the strong sector and their coefficients ($c_{R/Lv},\, c_{R/La},\, c_{R/Lr}\,\textrm{and}\, c_{Qq}^{(8,3)}$) scale like $g_\rho^0$ at best. On the other hand, the coefficient $c_{hg}$ associated with the operator $\mathcal{O}_{hg}$ scales as $g_\rho^{-1}$ (if only one field is composite), $g_\rho^{0}$ (if only two fields are composite) or $g_\rho$ (if the three fields are composite)
\end{itemize}
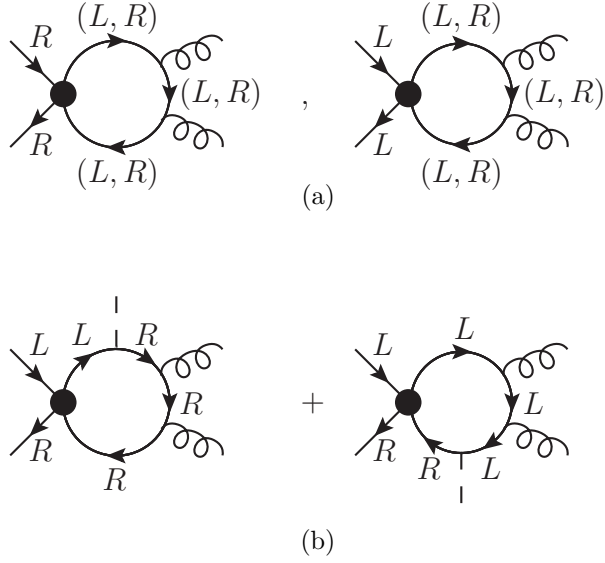
\begin{figure}[!bt]
	\begin{center}
\fcolorbox{white}{white}{
  \subfigure[(a)]{\begin{picture}(228,76) (19,9)
    \SetWidth{0.8}
    \SetColor{Black}
    \Line[arrow,arrowpos=0.5,arrowlength=5,arrowwidth=2,arrowinset=0.2](20,56)(40,36)
    \Line[arrow,arrowpos=0.5,arrowlength=5,arrowwidth=2,arrowinset=0.2](40,36)(20,16)
    \Arc[arrow,arrowpos=0.425,arrowlength=5,arrowwidth=2,arrowinset=0.2,flip](60,36)(20.125,117,477)
    \Arc[arrow,arrowpos=0.925,arrowlength=5,arrowwidth=2,arrowinset=0.2,flip](60,36)(20.125,117,477)
    \Arc[arrow,arrowpos=0.675,arrowlength=5,arrowwidth=2,arrowinset=0.2,flip](60,36)(20.125,117,477)
    \Vertex(40,36){5}
    \Gluon(77,46)(100,56){4}{2}
    \Gluon(77,26)(100,16){4}{2}
    \Line[arrow,arrowpos=0.5,arrowlength=5,arrowwidth=2,arrowinset=0.2](150,56)(170,36)
    \Line[arrow,arrowpos=0.5,arrowlength=5,arrowwidth=2,arrowinset=0.2](170,36)(150,16)
    \Arc[arrow,arrowpos=0.38,arrowlength=5,arrowwidth=2,arrowinset=0.2,flip](190,36)(19.105,133,493)
    \Arc[arrow,arrowpos=0.88,arrowlength=5,arrowwidth=2,arrowinset=0.2,flip](190,36)(19.105,133,493)
    \Arc[arrow,arrowpos=0.63,arrowlength=5,arrowwidth=2,arrowinset=0.2,flip](190,36)(19.105,133,493)
    \Vertex(170,36){5}
    \Gluon(206,46)(230,56){4}{2}
    \Gluon(206,26)(230,16){4}{2}
    \Text(27,54)[lb]{\normalsize{\Black{$R$}}}
    \Text(27,13)[lb]{\normalsize{\Black{$R$}}}
    \Text(157,54)[lb]{\normalsize{\Black{$L$}}}
    \Text(157,13)[lb]{\normalsize{\Black{$L$}}}
    \Text(175,59)[lb]{\normalsize{\Black{$(L,R)$}}}
    \Text(214,30)[lb]{\normalsize{\Black{$(L,R)$}}}
    \Text(175,0)[lb]{\normalsize{\Black{$(L,R)$}}}
    \Text(45,59)[lb]{\normalsize{\Black{$(L,R)$}}}
    \Text(84,30)[lb]{\normalsize{\Black{$(L,R)$}}}
    \Text(45,0)[lb]{\normalsize{\Black{$(L,R)$}}}
    \Text(130,30)[lb]{\normalsize{\Black{$,$}}}
  \end{picture}}
}\end{center}
\begin{center}
\fcolorbox{white}{white}{
  \subfigure[(b)]{\begin{picture}(228,88) (19,7)
    \SetWidth{0.8}
    \SetColor{Black}
    \Line[arrow,arrowpos=0.5,arrowlength=5,arrowwidth=2,arrowinset=0.2](20,68)(40,48)
    \Line[arrow,arrowpos=0.5,arrowlength=5,arrowwidth=2,arrowinset=0.2](40,48)(20,28)
    \Arc[arrow,arrowpos=0.425,arrowlength=5,arrowwidth=2,arrowinset=0.2,flip](60,48)(20.125,117,477)
    \Arc[arrow,arrowpos=0.825,arrowlength=5,arrowwidth=2,arrowinset=0.2,flip](60,48)(20.125,117,477)
    \Arc[arrow,arrowpos=1.025,arrowlength=5,arrowwidth=2,arrowinset=0.2,flip](60,48)(20.125,117,477)
    \Arc[arrow,arrowpos=0.675,arrowlength=5,arrowwidth=2,arrowinset=0.2,flip](60,48)(20.125,117,477)
    \Vertex(40,48){5}
    \Gluon(77,58)(100,68){4}{2}
    \Gluon(77,38)(100,28){4}{2}
    \Line[arrow,arrowpos=0.5,arrowlength=5,arrowwidth=2,arrowinset=0.2](150,68)(170,48)
    \Line[arrow,arrowpos=0.5,arrowlength=5,arrowwidth=2,arrowinset=0.2](170,48)(150,28)
    \Arc[arrow,arrowpos=0.88,arrowlength=5,arrowwidth=2,arrowinset=0.2,flip](190,48)(19.105,133,493)
    \Arc[arrow,arrowpos=1.28,arrowlength=5,arrowwidth=2,arrowinset=0.2,flip](190,48)(19.105,133,493)
    \Arc[arrow,arrowpos=1.48,arrowlength=5,arrowwidth=2,arrowinset=0.2,flip](190,48)(19.105,133,493)
    \Arc[arrow,arrowpos=0.63,arrowlength=5,arrowwidth=2,arrowinset=0.2,flip](190,48)(19.105,133,493)
    \Vertex(170,48){5}
    \Gluon(206,58)(230,68){4}{2}
    \Gluon(206,38)(230,28){4}{2}
    \Text(27,66)[lb]{\normalsize{\Black{$L$}}}
    \Text(27,26)[lb]{\normalsize{\Black{$R$}}}
    \Text(157,66)[lb]{\normalsize{\Black{$L$}}}
    \Text(157,26)[lb]{\normalsize{\Black{$R$}}}
    \Text(214,44)[lb]{\normalsize{\Black{$L$}}}
    \Text(84,44)[lb]{\normalsize{\Black{$R$}}}
    \Text(130,44)[lb]{\normalsize{\Black{$+$}}}
    \Line[dash,dashsize=5](60,69)(60,88)
    \Line[dash,dashsize=5](190,29)(190,10)
    \Text(67,70)[lb]{\normalsize{\Black{$R$}}}
    \Text(43,70)[lb]{\normalsize{\Black{$L$}}}
    \Text(55,15)[lb]{\normalsize{\Black{$R$}}}
    \Text(188,72)[lb]{\normalsize{\Black{$L$}}}
    \Text(198,19)[lb]{\normalsize{\Black{$L$}}}
    \Text(174,19)[lb]{\normalsize{\Black{$R$}}}
  \end{picture}}
}\end{center}
\caption{\small Typical one loop contributions of (a) the dimension-six operators (\ref{odr})--(\ref{odb}) leading to $\delta c_{Rv}$ and $\delta c_{Lv}$ respectively once the equation of motion~\eqref{motion} is used, and (b) the dimension-eight operator $\left(H\bar{Q}t\right)\left(H\bar{Q}t\right)$ leading to $\delta c_{hg}$ if one chirality-flip is considered in the loop.}
\label{Fig:oneloop}
\end{figure}
In the limit $g_\rho\sim4\pi$, the one-loop contributions of the dominant operators (\ref{odr})--(\ref{odb}) to top pair production may be as large as the tree-level contributions of the subdominant ones given in Section~\ref{sec:relop}. However, the chiral structure of the dominant operators are such that their one-loop corrections (see Fig.~\ref{Fig:oneloop}a and~\ref{Fig:oneloop}b) simply amount to redefining the coefficients $c_{Rv}$ and $c_{Lv}$ in the Lagrangian~\eqref{eq:ttlag}~\cite{Gerard:2004je}:
\begin{eqnarray}
\frac{\delta c_{Rv}}{g_s^2} &=&  \frac{{g_S^{(1)}}^2-{g_S^{(8)}}^2/6-4g_R^2}{3\left(4\pi\right)^2}\log\left(\frac{\Lambda^2}{m_t^2}\right) +\frac{{g_S^{(1)}}^2-{g_S^{(8)}}^2/6}{3\left(4\pi\right)^2}\log\left(\frac{\Lambda^2}{m_b^2}\right)\nonumber\\
\frac{\delta c_{Lv}}{g_s^2} &=&  \frac{{g_S^{(1)}}^2-{g_S^{(8)}}^2/6-4 {g_L^{(1)}}^2+8 {g^{(8)}_L}^2/3}{3\left(4\pi\right)^2} \log\left(\frac{\Lambda^2}{m_t^2}\right) + \frac{2 {g^{(8)}_L}^2}{3\left(4\pi\right)^2} \log\left(\frac{\Lambda^2}{m_b^2}\right)
\end{eqnarray}
where $g_R^2$, ${g_L^{(i)}}^2$ and ${g_S^{(i)}}^2$ are the coefficients of the operator ${\mathcal O}_R$, ${\mathcal O}_L^{(i)}$ and ${\mathcal O}_S^{(i)}$ respectively. 
The operator $\left(\bar{t}_L t_R\right)\left(\bar{t}_L t_R\right)$ and $\left(\bar{t}_R t_L\right)\left(\bar{t}_R t_L\right)$ would induce a modification of $c_{hg}$ at one loop~\cite{Gerard:2004je}. However, $SU(2)$ gauge invariance requires to consider loop corrections induced by a dimension-eight operator like $\left(H\bar{Q}t\right)\left(H\bar{Q}t\right)$ with the Higgs field $H$ replaced by its {\it vev} (see Fig.~\ref{Fig:oneloop}b). 
\begin{figure}
\begin{center}
\fcolorbox{white}{white}{
  \begin{picture}(289,76) (89,-219)
    \SetWidth{0.8}
    \SetColor{Black}
    \Photon(110,-194)(160,-196){4}{3}
    \Line[arrow,arrowpos=0.5,arrowlength=5,arrowwidth=2,arrowinset=0.2](160,-196)(180,-174)
    \Line[arrow,arrowpos=0.5,arrowlength=5,arrowwidth=2,arrowinset=0.2](180,-214)(160,-196)
    \Text(135,-188)[lb]{\normalsize{\Black{$V$}}}
    \Text(75,-172)[lb]{\normalsize{\Black{$(L,R)$}}}
    \Text(75,-230)[lb]{\normalsize{\Black{$(L,R)$}}}
    \Text(165,-172)[lb]{\normalsize{\Black{$(L,R)$}}}
    \Text(165,-230)[lb]{\normalsize{\Black{$(L,R)$}}}
    \Text(130,-245)[lb]{\normalsize{\Black{$(a)$}}}
    \Line[arrow,arrowpos=0.5,arrowlength=5,arrowwidth=2,arrowinset=0.2](90,-174)(110,-194)
    \Line[arrow,arrowpos=0.5,arrowlength=5,arrowwidth=2,arrowinset=0.2](110,-194)(90,-214)
    \Line[arrow,arrowpos=0.5,arrowlength=5,arrowwidth=2,arrowinset=0.2](110,-194)(90,-214)
    \Line[arrow,arrowpos=0.5,arrowlength=5,arrowwidth=2,arrowinset=0.2](160,-196)(180,-174)
    \Line[arrow,arrowpos=0.5,arrowlength=5,arrowwidth=2,arrowinset=0.2](90,-174)(110,-194)
    \Line[arrow,arrowpos=0.5,arrowlength=5,arrowwidth=2,arrowinset=0.2](270,-174)(290,-194)
    \Line[arrow,arrowpos=0.5,arrowlength=5,arrowwidth=2,arrowinset=0.2](290,-194)(270,-214)
    \Photon(290,-194)(320,-194){4}{2}
    \Arc[arrow,arrowpos=0.425,arrowlength=5,arrowwidth=2,arrowinset=0.2,flip](340,-194)(20.125,117,477)
    \Arc[arrow,arrowpos=0.825,arrowlength=5,arrowwidth=2,arrowinset=0.2,flip](340,-194)(20.125,117,477)
    \Arc[arrow,arrowpos=1.025,arrowlength=5,arrowwidth=2,arrowinset=0.2,flip](340,-194)(20.125,117,477)
    \Arc[arrow,arrowpos=0.675,arrowlength=5,arrowwidth=2,arrowinset=0.2,flip](340,-194)(20.125,117,477)
    \Gluon(358,-185)(380,-173){4}{2}
    \Gluon(358,-204)(380,-213){4}{2}
    \Line[dash,dashsize=5](340,-174)(340,-154)
    \Text(293,-188)[lb]{\normalsize{\Black{$S(T)$}}}
    \Text(277,-178)[lb]{\normalsize{\Black{$R$}}}
    \Text(277,-216)[lb]{\normalsize{\Black{$L$}}}
    \Text(323,-174)[lb]{\normalsize{\Black{$L$}}}
    \Text(350,-174)[lb]{\normalsize{\Black{$R$}}}
    \Text(363,-199)[lb]{\normalsize{\Black{$R$}}}
    \Text(336,-229)[lb]{\normalsize{\Black{$R$}}}
    \Text(326,-245)[lb]{\normalsize{\Black{$(b)$}}}
  \end{picture}
}\end{center}\vspace{3mm}
\caption{\small One particle exchange contributions to $\mathcal{L}_{t\bar{t}}$ in Eq.~\eqref{eq:ttlag}: (a) the five four-fermion operators can be directly associated with the exchange of a spin-1 resonance once Fierz transformations are used, (b) the single two-fermion operator $\mathcal{O}_{hg}$ can be indirectly associated with the exchange of a spin-0 or spin-2 resonance coupled to two gluons via a fermion loop.}\label{Fig:reso}
\end{figure}
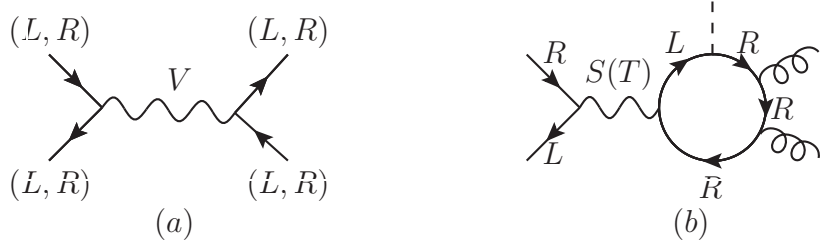

Similarly, the connection within resonance models is quite straightforward using the standard Fierz relations of App.~\ref{app:Fierz}. The exchanges of heavy vectors and scalars lead to four-fermion operators (explicit formulas are given for instance in Ref.~\cite{Jung:2009pi}) but cannot contribute to the top chromo-magnetic moment at tree-level as a consequence of $SU(3)_c$ gauge invariance (see Fig.~\ref{Fig:reso}a).
Only higher-dimension effective operators quadratic in the gluon field-strength can be induced in this frame. For example, a heavy scalar or tensor induces at tree-level the operator $\left(H\bar{Q}t+h.c.\right)G_{\mu\nu}G^{\mu\nu}$ or $\left(H\bar{Q}t-h.c.\right)G_{\mu\nu}\tilde{G}^{\mu\nu}$ (see Fig.~\ref{Fig:reso}b). So, the operator $\mathcal{O}_{hg}$ can only be generated at the loop-level and is suppressed in resonance models. 

Finally, note that a generic prediction of recent
models of Higgs and top compositeness
is the existence of fermionic top partners (``custodians") whose mass is below the mass scale of the vector resonances $m_{\rho}$ \cite{Contino:2006qr,Contino:2008hi,Mrazek:2009yu}.
They do not appear in our low-energy effective field theory approach. However, they could be produced singly or in pairs at the LHC, and decay either into $tZ$, $tH$, $bW$ or $tW$,  therefore possibly making a new physics ``background" to $t\bar{t}+X$ final states. We leave the study of these effects and their description in terms of effective operators to future investigations.

\section{Top pair production cross section}
\label{sec:ttbar}

We now move to the core of our model-independent analysis whose goal is to evaluate the LHC potential in probing new physics in top pair production beyond  the Tevatron's reach.

\subsection{Partonic differential cross~sections}\label{sec:partonic}

As already mentioned, top pair production is calculated at the same order in ${1}/{\Lambda}$ as the Lagrangian in Eq.~\eqref{eq:ttlag}
\begin{equation}
\left|M\right|^2 = \left|M_{SM}\right|^2 + 2 \Re{\left(M_{SM} M_{NP}^*\right)}+ \mathscr{O}\left(\frac{1}{\Lambda^4}\right),
\end{equation}
where $M_{NP}$ represents the matrix element of all the (new physics) dimension-six operators introduced in Section~\ref{sec:EffLag}. 

From the Lagrangian in Eq.~\eqref{eq:ttlag}, the two parton-level cross~sections for $t\bar{t}$ production at $\mathscr{O}\left(\Lambda^{-2}\right)$ follow from the Feynman diagrams depicted in Fig.~\ref{Fig:gg} and~\ref{Fig:qq} of App.~\ref{app:Feynman}. Their expressions are ($v=246$~GeV): 
\begin{eqnarray}
\!\!\!&\!\!\!&\!\!\!\frac{d\sigma}{dt}\left(q\bar{q}\rightarrow t\bar{t}\right) = \frac{d\sigma_{SM}}{dt}\left(1+\frac{c_{Vv}\pm \frac{c_{Vv}^\prime}{2}}{g_s^2}\frac{s}{\Lambda^2}\right)+\frac{1}{\Lambda^2}\frac{\alpha_s}{9s^2} \left(\left(c_{Aa}\pm \frac{c_{Aa}^\prime}{2}\right) s(\tau_2-\tau_1)+4 g_s c_{hg}\sqrt{2}vm_t\right)\nonumber\\\!\!\!&\!\!\!&\!\!\!\label{qq}\\
\!\!\!&\!\!\!&\!\!\!\frac{d\sigma}{dt}\left(gg\rightarrow t\bar{t}\right) = \frac{d\sigma_{SM}}{dt} +\sqrt{2} \alpha_s g_s\frac{vm_t}{s^2} \frac{c_{hg}}{\Lambda^2} \left(\frac{1}{6\tau_1\tau_2}-\frac{3}{8}\right)\label{gg}
\end{eqnarray}
where the upper (lower) sign is for the up (down) quarks and 
\begin{eqnarray}
\!\!\!&\!\!\!&\!\!\!\frac{d\sigma_{SM}}{dt}\left(q\bar{q}\rightarrow t\bar{t}\right) = \frac{4 \pi \alpha_s ^2}{9s^2} \left(\tau_1^2+\tau_2^2+\frac{\rho}{2}\right) \label{SMqq}\\
\!\!\!&\!\!\!&\!\!\!\frac{d\sigma_{SM}}{dt}\left(gg\rightarrow t\bar{t}\right) = \frac{\pi \alpha_s^2}{s^2} \left(\frac{1}{6\tau_1\tau_2}-\frac{3}{8}\right)(\rho+\tau_1^2+\tau_2^2-\frac{\rho^2}{4\tau_1\tau_2})\qquad\qquad\qquad\qquad\qquad\qquad\qquad\label{SMgg}
\end{eqnarray}
\begin{equation}
\mbox{with} \ \   \tau_1=\frac{m_t^2-t}{s},\quad \tau_2=\frac{m_t^2-u}{s}, \quad \rho=\frac{4m_t^2}{s}.
\end{equation}
 The Mandelstam parameter $t$ is related,  in the $t\bar{t}$ center-of-mass frame,  to the angle $\theta$ between the momenta of the incoming parton and the outgoing top quark by ($\beta=\sqrt{1-\frac{4m^2}{s}}$)
\begin{equation}
m_t^2-t=  \frac{s}{2} \left( 1- \beta\cos \theta \right).
\label{tcosthetarelation}
\end{equation}
All the contributions to the $t\bar{t}$ differential cross~section but the one proportional to $c_{Aa}\pm \frac{c_{Aa}^\prime}{2}$ are invariant under $\theta \to \pi -\theta$. 

Similar results have already been derived in the literature.
For instance, these cross~sections were recently fully computed in Ref.~\cite{Zhang:2010dr} and consistent  with our expressions with the identifications given in Table~\ref{tab:comparison}. This non exhaustive table also gives the correspondences with respect to some other recent works ~\cite{Kumar:2009vs,Cao:2010zb,Jung:2009pi,Hioki:2009hm}. Note that the contribution of the chromomagnetic operator $\mathcal{O}_{hg}$ has been extensively discussed in the literature~\cite{Atwood:1994vm,Cheung:1995nt,Whisnant:1997qu,Hikasa:1998wx} and recently revisited for both processes in Ref.~\cite{Hioki:2009hm,Choudhury:2009wd}.
%
%

%
%
\begin{table}[!htb]
\centering{
\small
\begin{tabular}{|l|c|c|c|c|c|}
\hline
&Ref.~\cite{Zhang:2010dr}&Ref.~\cite{Kumar:2009vs}&Ref.~\cite{Cao:2010zb}&Ref.~\cite{Jung:2009pi}&Ref.~\cite{Hioki:2009hm}\\
\hline
{\vrule height 12pt depth 0pt width 0pt}
$c_{hg}$&$2C_{tG}$&$g_1 g_s$&&&$\frac{1}{2}C^{33}_{uG\phi}$\\[.1cm]
\hline
{\vrule height 15pt depth 0pt width 0pt} 
$c_{Vv}$&$\frac{1}{4}\left(C^1_{u}+C^2_{u}+C^1_{d}+C^2_{d}\right)$&$-g_2 g_s^2$(*)&$\frac{g_s^2}{4}(\kappa_R^u+\kappa_R^d+\kappa_L^u+\kappa_L^d)$(*)&$\frac{g_s^2}{2}(C_1+C_2)$& \\[.1cm]
\hline
{\vrule height 15pt depth 0pt width 0pt} 
$c_{Aa}$&$\frac{1}{4}\left(C^1_{u}-C^2_{u}+C^1_{d}-C^2_{d}\right)$& &$\frac{g_s^2}{4}(\kappa_R^u+\kappa_R^d+\kappa_L^u+\kappa_L^d)$(*)&$\frac{g_s^2}{2}(C_1-C_2)$& \\[.1cm]
\hline
{\vrule height 15pt depth 0pt width 0pt} 
$c_{Vv}^\prime$&$\frac{1}{2}\left(C^1_{u}+C^2_{u}-C^1_{d}-C^2_{d}\right)$& & $\frac{g_s^2}{2}(\kappa_R^u-\kappa_R^d+\kappa_L^u-\kappa_L^d)$(*)& & \\[.1cm]
\hline
{\vrule height 15pt depth 0pt width 0pt} 
$c_{Aa}^\prime$&$\frac{1}{2}\left(C^1_{u}-C^2_{u}-C^1_{d}+C^2_{d}\right)$& &$\frac{g_s^2}{2}(\kappa_R^u-\kappa_R^d+\kappa_L^u-\kappa_L^d)$(*) & & \\[0.1cm]
\hline
 \end{tabular}
\caption{\small Dictionary between our parameters and those used in recent papers on the subject. They all agree eventually up to a sign for those that are labeled by a (*). For Ref.~\cite{Zhang:2010dr}, $C_{qq}^{(8,3)}={c_{Qq}^{(8,3)}}$. Blank entries mean that the corresponding operators were not considered. \label{tab:comparison} }
}
\end{table}

As can be seen from Eqs.~\eqref{SMgg} and~\eqref{gg}, the new physics and the SM contributions for gluon fusion have a common factor. In fact, this common factor is what is mainly responsible for the shape of the distributions of the SM. This is the reason why, as we will stress again in the following, the operator $\mathcal{O}_{hg}$ can hardly be distinguished from the SM in gluon fusion. 

Equation~\eqref{qq} shows that only two kinds of four-fermion operators actually contribute to the differential cross-section after averaging over the final state spins:
\begin{itemize}
	\item the first one is responsible for the even part in the scattering angle proportional to $c_{Vv}\pm\frac{c_{Vv}^\prime}{2}$
	\begin{equation}
	\bar{t}\gamma^\mu T^A t \bar{q} \gamma^\mu T^A q
	\end{equation}
	where here $t$ and $q=u,d$ stand for the full 4-component Dirac spinor; 
	\item the second one is responsible for the odd part in the scattering angle proportional to $c_{Aa}\pm\frac{c_{Aa}^\prime}{2}$
	\begin{equation}
	\bar{t}\gamma^\mu\gamma_5 T^A t \bar{q} \gamma^\mu\gamma_5 T^A q.
	\end{equation}
\end{itemize}

\subsection{Total cross~section}
\label{sec:totalcs}

\subsubsection{LHC--Tevatron complementarity}

Since the dependence on $c_{Aa}$ and $c'_{Aa}$ vanishes after the integration over the kinematical variable $t$, the  total cross~section depends thus only on the three parameters $c_{hg}$ $c_{Vv}$ and $c'_{Vv}$. Moreover, the $t\bar{t}$ production by gluon fusion only depends on the coefficient of the operator ${\mathcal O}_{hg}$. 
Our results for $t\bar{t}$ production are obtained by the convolution of the analytic differential cross~section of Eqs.~\eqref{qq} 
and~\eqref{gg} with the pdf (taking  CTEQ6L1~\cite{Nadolsky:2008zw}). We have also implemented the new vertices in MadGraph~\cite{Alwall:2007st} and used them to validate our results. 
At leading order, we have \\[2mm]
--- at the LHC ($\sqrt{s}=14$~TeV):
\begin{eqnarray}
\label{resultLO1}
\sigma\left(gg\rightarrow t\bar{t}\right)/{\mbox{pb}} &\!\! = \!\! & 466^{+146}_{-103}+\left(127_{-23}^{+31}\right) c_{hg}
\left(\frac{1\mbox{ TeV}}{\Lambda} \right)^2\!\!,\\
\label{resultLO2}
\sigma\left(q\bar{q}\rightarrow t\bar{t}\right)/\mbox{pb} &\!\! = \!\! & 72^{+16}_{-12}+\left[
\left(15_{-1}^{+2}\right) c_{Vv}  +\left(17_{-2}^{+3}\right) c_{hg}  +\left(1.32_{-0.12}^{+0.12}\right) c_{Vv}^\prime \right]\!\!\left(\frac{1\mbox{ TeV}}{\Lambda} \right)^2\!\!,\\
\label{resultLO3}
\sigma\left(p p\rightarrow t\bar{t}\right)/\mbox{pb} &\!\! = \!\! & 538^{+162}_{-115}+\left[
\left(15_{-1}^{+2}\right) c_{Vv}  +\left(144_{-25}^{+34}\right) c_{hg} +\left(1.32_{-0.12}^{+0.12}\right) c_{Vv}^\prime \right] \!\!\left(\frac{1\mbox{ TeV}}{\Lambda} \right)^2\!\!.\qquad\quad\,\,\,
\end{eqnarray}
--- at the LHC ($\sqrt{s}=7$~TeV):
\begin{eqnarray}
 \sigma\left(p p\rightarrow t\bar{t}\right)/\mbox{pb} &\!\!  = \!\!  & 94^{+22}_{-17}+\left[ \left(4.5^{+0.7}_{-0.6}\right) c_{Vv} + \left(25^{+7}_{-5}\right) c_{hg} +\left(0.48^{+0.068}_{-0.056} \right)c_{Vv}^\prime \right]
  \!\!\left(\frac{1\mbox{ TeV}}{\Lambda} \right)^2\!\!.\qquad\quad\,\,\,
\end{eqnarray}
--- at the Tevatron ($\sqrt{s}=1.96$~TeV):
\begin{eqnarray}
\label{resultLO1b}
\sigma\left(gg\rightarrow t\bar{t}\right) /\mbox{pb} &\!\!\! = \!\!\! &0.35^{+0.20}_{-0.12}+\left(0.10_{-0.03}^{+0.05}\right) c_{hg}
 \left(\frac{1\mbox{ TeV}}{\Lambda} \right)^2\!\!,\\
 \label{resultLO2b}
\sigma\left(q\bar{q}\rightarrow t\bar{t}\right) /\mbox{pb} &\!\!\! = \!\!\! & 5.80^{+2.21}_{-1.49}+\!\!\left[\left(0.87_{-0.16}^{+0.23}\right) c_{Vv}+\left(1.34_{-0.30}^{+0.42}\right) c_{hg}  +\left(0.31_{-0.06}^{+0.08}\right) c_{Vv}^\prime \right]\!\!\left(\frac{1\mbox{ TeV}}{\Lambda} \right)^2\!\!,\nonumber\\ 
\quad\\
\label{resultLO3b}
\sigma\left(p p\rightarrow t\bar{t}\right) /\mbox{pb} &\!\!\! = \!\!\! & 6.15^{+2.41}_{-1.61}+\!\!\left[
\left(0.87_{-0.16}^{+0.23}\right) c_{Vv} +\left(1.44_{-0.33}^{+0.47}\right) c_{hg} +\left(0.31_{-0.06}^{+0.08}\right) c_{Vv}^\prime \right]\!\!\left(\frac{1\mbox{ TeV}}{\Lambda} \right)^2\!\!. \nonumber\\ 
\quad
\end{eqnarray}
Numerically, the contribution from the isospin-1 sector ($c_{Vv}^\prime$) is suppressed compared to the contribution of the isospin-0 sector ($c_{Vv}$) and this suppression is more effective at the LHC than at the Tevatron. This is due to the fact that, at Tevatron, the top pair production by up-quark annihilation is $5\div 6$ times bigger than by down-quark annihilation. At the LHC, this ratio is reduced to 1.4 only. In most of the rest of our analysis, we shall neglect the contribution from the isospin-1 sector since it is subdominant.

\begin{figure}[!ht]
	\centering
		\includegraphics[width=0.66\textwidth]{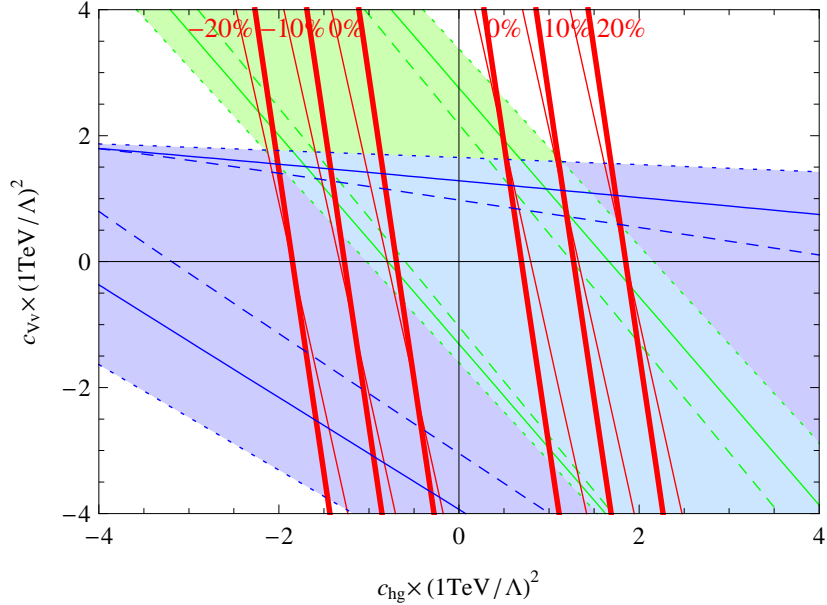}
		\caption{	\label{Mttconstraints}
\small Region allowed by the Tevatron constraints (at 2$\sigma$) for $c'_{Vv}=0$. The green region is allowed by the total cross~section measurement.  The blue region is  consistent with the $t \bar t$ invariant mass shape. 	The red lines show the limits that can be set by the LHC at 7~TeV (thin line) and at 14~TeV (thick line) as soon as a precision on the top pair cross~section of 10\% and 20\% respectively is reached. The ``$0 \%$"line delimits the region where the new physics contributions are smaller than the theoretical error on the SM cross~section. The dashed  ($\mu_F=\mu_R=\frac{m_t}{2}$), dotted ($\mu_F=\mu_R=2 m_t$) and solid lines ($\mu_F=\mu_R=m_t=174.3$ GeV) show the estimated theoretical uncertainties.}
\end{figure}

The measurements of the total cross~section at the Tevatron and at the LHC are complementary as shown in Fig.~\ref{Mttconstraints}.  As expected, the LHC $pp\to t\bar{t}$ total cross~section strongly depends on $c_{hg}$. Consequently, it can be used to constrain directly the allowed range for  $c_{hg}$. On the contrary, the corresponding Tevatron cross-section depends on both  $c_{hg}$ and  $c_{Vv}$ and constrains thus a combination of these parameters.

\begin{figure}[!ht]
	\centering
		\includegraphics[width=0.495\textwidth]{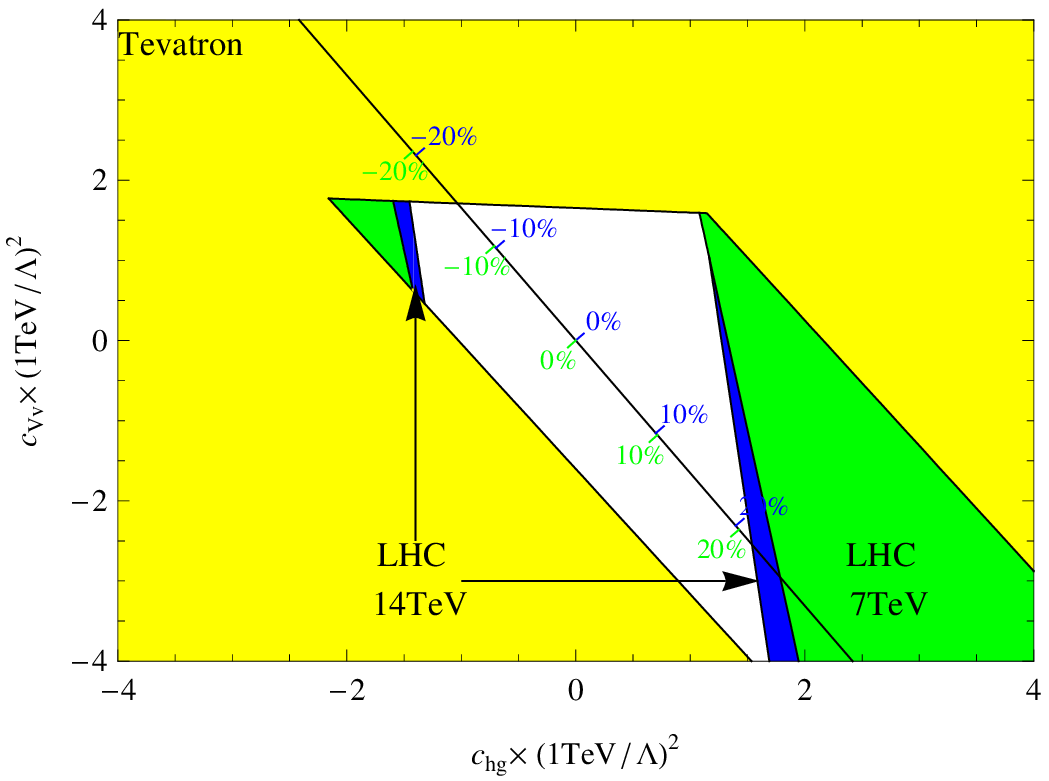}
		\includegraphics[width=0.495\textwidth]{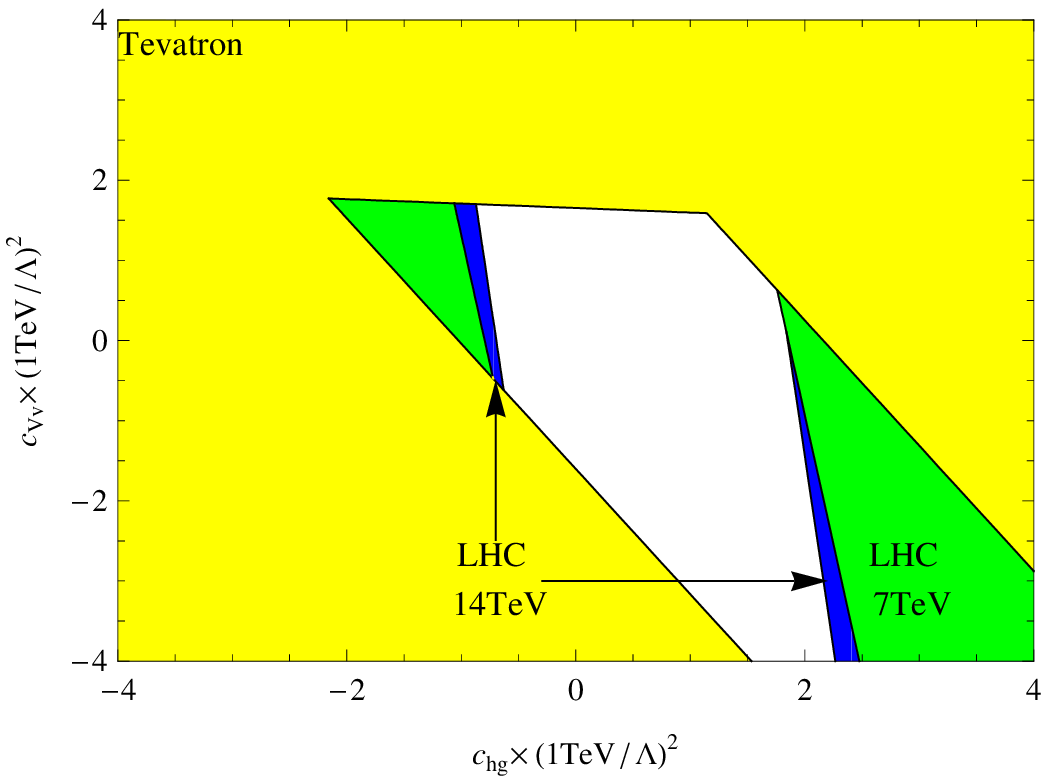}
		\includegraphics[width=0.495\textwidth]{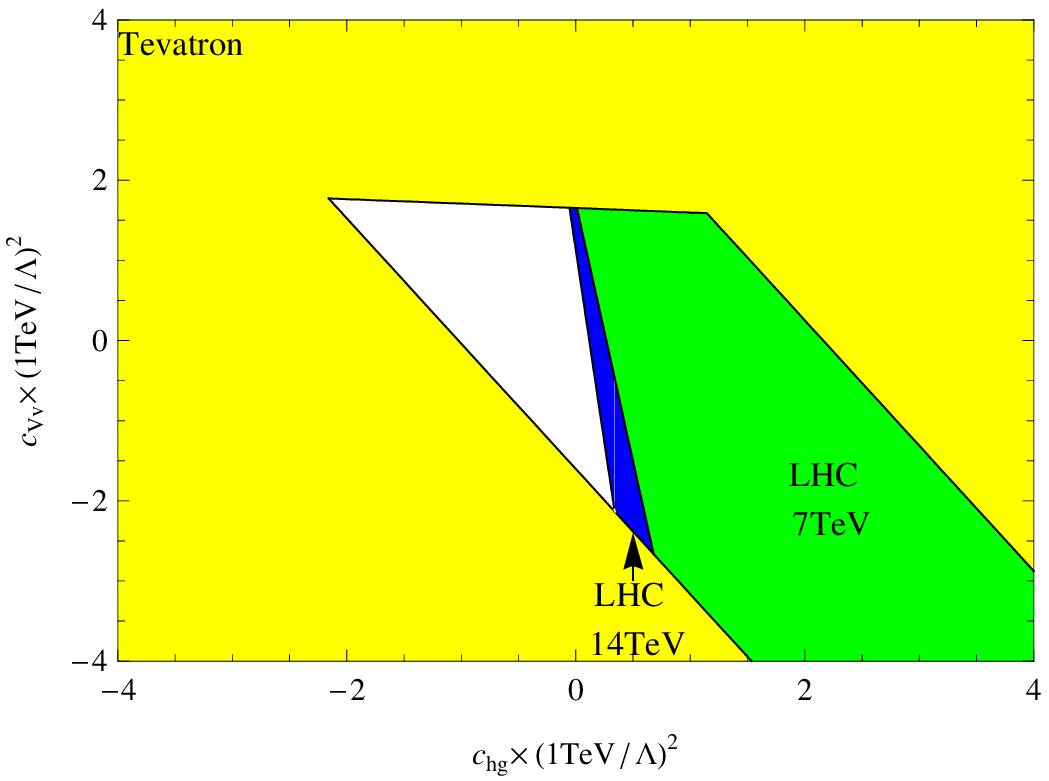}
		\caption{\label{SummaryPlot}
\small
Summary plot (taking $\mu_F=\mu_R=m_t$ and defining the exclusion region at $2\sigma$). The yellow region is excluded by the Tevatron. The green (blue) region is  excluded by LHC at 7~TeV (14~TeV) after a precision of 10\% is reached on the $t\bar t $ cross~section. In the first plot, it is assumed that the measured cross~section is the SM value. If the measured value deviates from the SM one, the center of the unconstrained white region will be translated along the thin black  line as indicated. For instance, the second and third plots show the displacement of the white region  if the measured $\sigma_{t\bar{t}}$ is respectively $+10\%$  and $-20\%$ of the SM value. }
\end{figure}

In Fig.~\ref{Mttconstraints}, we assume that the central measured value of the cross~section at the LHC coincides with the SM theoretical prediction. Figure~\ref{SummaryPlot} shows how the allowed region is shifted to the left (right) if the measured value is lower (higher) than the computed value. We use the NLO+NLL prediction~\cite{Cacciari:2008zb} for the SM cross~section ($m_t=174.3$~GeV) at the LHC
\begin{eqnarray}
\sigma^{\textrm{14 TeV}}_{\textrm{th}}&=&832^{+75}_{-78}(\textrm{scale})^{+28}_{-27}(\textrm{pdf})~\textrm{pb},\nonumber\\
\sigma^{\textrm{7 TeV}}_{\textrm{th}}&=&146^{+12}_{-13}(\textrm{scale})^{+11}_{-11}(\textrm{pdf})~\textrm{pb},
\end{eqnarray} 
and at the Tevatron
\begin{equation}
\sigma^{\textrm{1.96 TeV}}_{\textrm{th}}=6.87^{+0.26}_{-0.48}(\textrm{scale})^{+0.47}_{-0.33}(\textrm{pdf})~\textrm{pb}.\label{xst}
\end{equation} 
In Fig.~\ref{Mttconstraints}, we combine the errors linearly. For the experimental value, we use the CDF combination of all channel at 4.6 $\textrm{fb}^{-1}$~\cite{Cerrito:2010sv},
\begin{equation}
 \sigma^{\textrm{1.96 TeV}}_{\textrm{obs}} = 7.5 \pm 0.31 (\textrm{stat}) \pm 0.34 (\textrm{syst}) \pm 0.15 (\textrm{lumi})~\textrm{pb}
\end{equation} 
and combine the errors quadratically. Due to the rather large uncertainties on the theoretical normalization, the region allowed by the total cross~section measurement remains large. Even if the experimental precision becomes very good, a rather large allowed region will remain due to the theoretical uncertainties. An improvement of the theoretical prediction for top pair production in SM is necessary to reduce the allowed region. The theoretical uncertainties for the new physics part are estimated by changing the factorisation scale $\mu_F$ and the renormalisation scale $\mu_R$. The errors from the pdf are not computed. The errors on the exclusion regions at the LHC are not shown but are about 20\% and are symmetric (10\% on each side of the allowed region). 

The absence of a large deviation in the measurement of the cross~section at the Tevatron implies
$c_{Vv}\approx -1.6 \,c_{hg}$ if the scale of new physics is rather low. From the discussion at the end of the classification of Section~\ref{sec:Composite}, it would mean that $c_{hg}$ and $c_{Vv}$ are both of the ${\mathscr O}(g_\rho^0)$, indicating that either both chiralities of the top or one chirality of the top and the Higgs boson are composite fields. Compared to the SM prediction, this would give a maximum deviation of the order of 25\% for the $t\bar{t}$ production cross-section at the LHC when $c_{hg }\left(\frac{1 {\textrm{ TeV}}}{\Lambda}\right)^2 \sim 2$.

\subsubsection{Domain of validity of the results}
\label{sec:validity}

Our calculation is performed at order $\mathscr{O}(1/\Lambda^2)$ as we keep only the interference term between the dimension-six and the Standard Model and we neglect any contribution suppressed by higher power of $\Lambda$. 
The validity of our results is thus limited to values of new coupling parameters and $\Lambda$ satisfying
\begin{equation}
\left. \sigma \right|_{{\mathscr O}(\Lambda^{-2})} \gtrsim \ \kappa \  \left. \sigma \right|_{{\mathscr O}(\Lambda^{-n})}
\end{equation}
where $n>2$ and $\kappa$ should be at least~2 in order to keep higher order the correction below 50\%. We have estimated the size of the $\mathscr{O}(\Lambda^{-4})$ contributions by computing the squared amplitudes of each dimension-six operators with MadGraph and we find at the LHC$_{14\textrm{ TeV}}$:
\begin{equation}
\left.\sigma \right|_{{\mathscr O}(\Lambda^{-4})}\sim \sigma_{NP^2} = \left(22.5 \,c_{hg}^2+3.7 \,c_{Vv}^2 \right)\times \left(\frac{\mbox{1 TeV}}{\Lambda}\right)^4 \ \textrm{pb}
\end{equation}
and, at the Tevatron,
\begin{equation}
\left.\sigma \right|_{{\mathscr O}(\Lambda^{-4})}\sim \sigma_{NP^2} = \left(0.103 \,c_{hg}^2+0.060 \,c_{Vv}^2 \right)\times \left(\frac{\mbox{1 TeV}}{\Lambda}\right)^4 \ \textrm{pb}
\end{equation}
Therefore, at the Tevatron, our results apply to a region of parameter space bounded by $\left| c_i\right| \left(\frac{1 {\textrm{ TeV}}}{\Lambda}\right)^2\lesssim 14/\kappa$. At the LHC,  since the center-of-mass energy is larger, the reliable region shrinks to $\left|c_{hg}\right|\left(\frac{1 {\textrm{ TeV}}}{\Lambda}\right)^2\lesssim 6/\kappa$ and $\left|c_{Vv}\right|\left(\frac{1 {\textrm{ TeV}}}{\Lambda}\right)^2\lesssim 4/\kappa$. Nevertheless, outside this region, the effect of the new physics should remain more or less of the same order excepted of course if there is some huge cancellation. Moreover, the cross-section is expected to be harder and harder as operators of higher dimensions are included in the effective Lagrangian. Ultimately some resonance threshold will be reached, leading to a radically different cross-section than the one predicted by the Standard Model.

It was found recently in Ref.~\cite{AguilarSaavedra:2010zi} that for the four-fermion operators, there are $\mathscr{O}(1/\Lambda^4)$ corrections from non-interfering contributions that can be almost as large as the $\mathscr{O}(1/\Lambda^2)$ interfering contributions at the LHC if $\Lambda \sim$ 1 TeV. However, at the LHC, these four-fermion operators give small contributions compared to the chromomagnetic operator. So we can conclude that including non-interfering four-fermion operators  will not change much our numerical analysis.

\begin{figure}[!ht]
	\centering
		\includegraphics[width=0.5\textwidth]{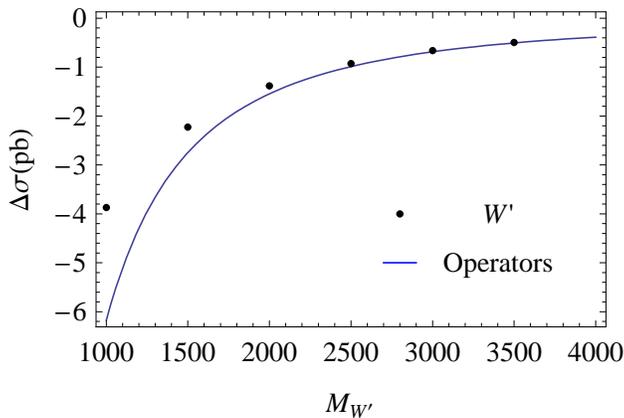}
		\caption{\label{fig:Comparison}
\small Correction to the SM cross-section at the LHC due to  a $W'$ and comparison with the effective field theory approach.
 }
\end{figure}

 Finally, to have an idea on how heavy the particles associated with new physics should be to allow an effective field theory treatment at the LHC, we compare in Fig.~\ref{fig:Comparison}  the correction to the SM cross-section at the LHC due to a $W^{\prime}$ (whose coupling to d and t quarks is 1) and the correction due to the corresponding effective operators ($C_{Vv}=-1/2$, $C'_{Vv}=-1$, $\Lambda=M_{W'}$). This plot shows that for $M_{W'}\gtrsim 1.5$ TeV the effective operators are a very good approximation (up to a few percents) at the LHC, although this depends  on the coupling.
 We will show in Fig.~\ref{fig:Comparison2} that a similar conclusion is reached at the Tevatron. 
 Consequently, the resonance models cannot be constrained in our effective approach since the exclusion regions in Fig.~\ref{SummaryPlot} correspond, for example, to a relatively light resonance ($M \lesssim$ TeV) with a coupling of order 1.
%

\subsection{$t\bar{t}$ invariant-mass, $p_T$ and $\eta$ distributions}\label{sec:distributions}

It was shown in Ref.~\cite{Kumar:2009vs} that the operators  ${\mathcal O}_{hg}$ and ${\mathcal O}_{Rv}$ can modify the invariant mass distribution at the Tevatron without drastically affecting the total cross~section, although no constraint was derived explicitly. We use in this section the latest CDF data~\cite{Goldschmidt:2010} to further constrain new physics. See also Ref.~\cite{Cao:2010zb} for a similar study on the $\bar{L}L\bar{L}L$ and $\bar{R}R\bar{R}R$ operators with the first data~\cite{Aaltonen:2009iz}. Since we have already used the measured total cross~section to constrain the parameter space here we only employ the shape information. 

For the sake of simplicity, in our analysis we assume that the measured values $m_i$ are normally distributed around the corresponding theoretical predictions $t_i$ with a standard deviation $\sigma_i$ given by their errors. Errors coming from different sources have been combined quadratically. We multiply by a common free coefficient $\zeta$ the theoretical prediction to get rid of the normalisation constraint. In practice, we use the best value for $\zeta$. The quantity 
\begin{equation}
\sum_{i=1}^n \frac{\left(m_i-\zeta t_i\right)^2}{\sigma_i^2}
\end{equation}
is then distributed as a $\chi^2$ with $n-1$ degrees of freedom. The theoretical predictions are obtained by integrating Eqs.~(\ref{qq}) and~(\ref{gg}) over the scattering angle. The explicit formulas are given in App.~\ref{app:xsec}. The SM distribution is computed at the tree level and normalised to the NLO+NLL result. The errors on the contribution of the operators are estimated by changing the factorisation and renormalisation scales. We take into account the bins between 350 GeV and 600 GeV ($n=13$). We cannot use the full distribution since our calculation only makes sense if $|g_{NP}| \frac{s}{\Lambda^2} \ll 1$. So $m_{t\bar{t}}\lesssim 1$ TeV if $\Lambda \sim 1$ TeV. The bound $m_{t\bar{t}}<600$~GeV seems reasonable since, even in the region $|g_{NP}| (\frac{\textrm{1 TeV}}{\Lambda})^2 \sim 4$, the estimation of the $1/\Lambda^4$ corrections from $\left|M_{NP}\right|^2$ are a bit less than $50\%$ of the $1/\Lambda^2$ corrections. For the next bins, these next order corrections become too large.

In Fig.~\ref{Mttconstraints}, we show the region consistent at 95\% C.L. with the $t \bar{t}$ invariant mass constraints reported in Ref.~\cite{Goldschmidt:2010}. As expected, the invariant mass shape is sensitive to a very different combination of the parameters than the total cross~section. Indeed, the interferences with the operators $\mathcal{O}_{Rv}$ and $\mathcal{O}_{Lv}$ grow faster than the SM by a factor $s$, which is not the case for $\mathcal{O}_{hg}$. The shape depends thus strongly on $c_{Vv}$. The Tevatron measurement already excludes the region $c_{Vv}\left(\frac{1\textrm{ TeV}}{\Lambda}\right)^2\gtrsim+2$.

\begin{figure*}[!htb]
 \centering
	\includegraphics[width=0.48\textwidth]{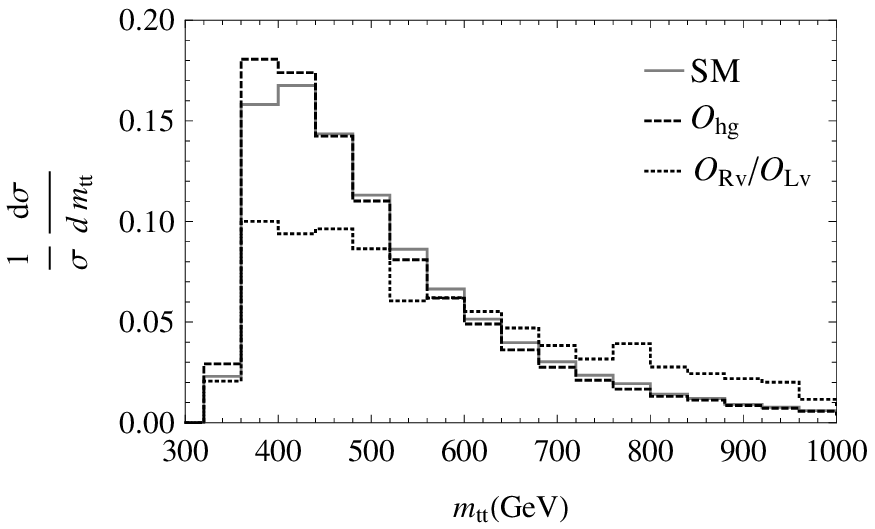}
	\includegraphics[width=0.48\textwidth]{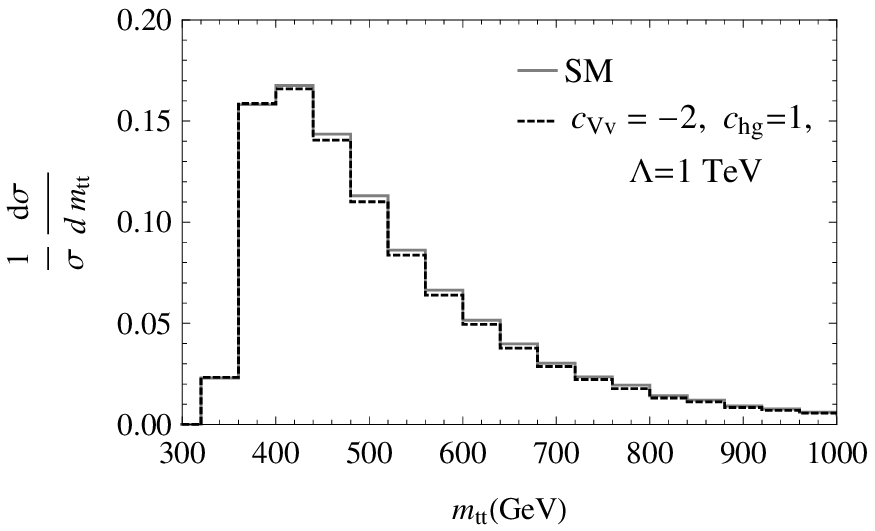}	
	\includegraphics[width=0.48\textwidth]{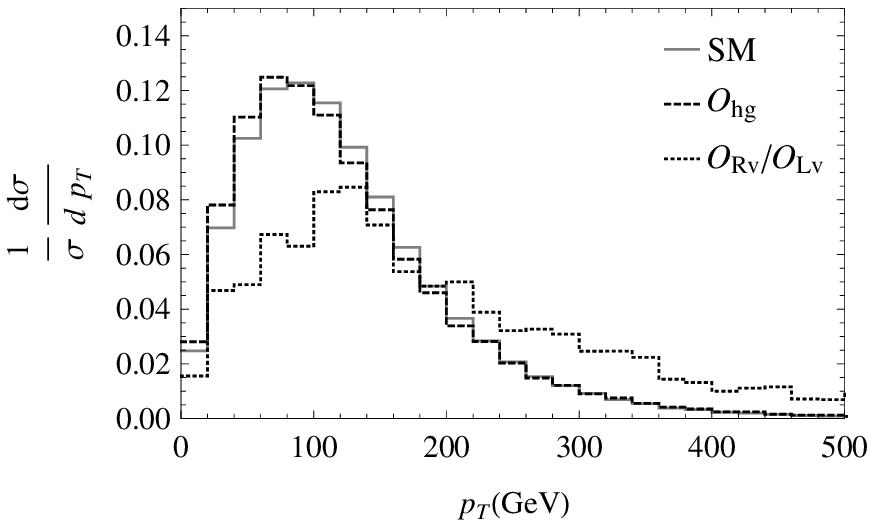}
	\includegraphics[width=0.48\textwidth]{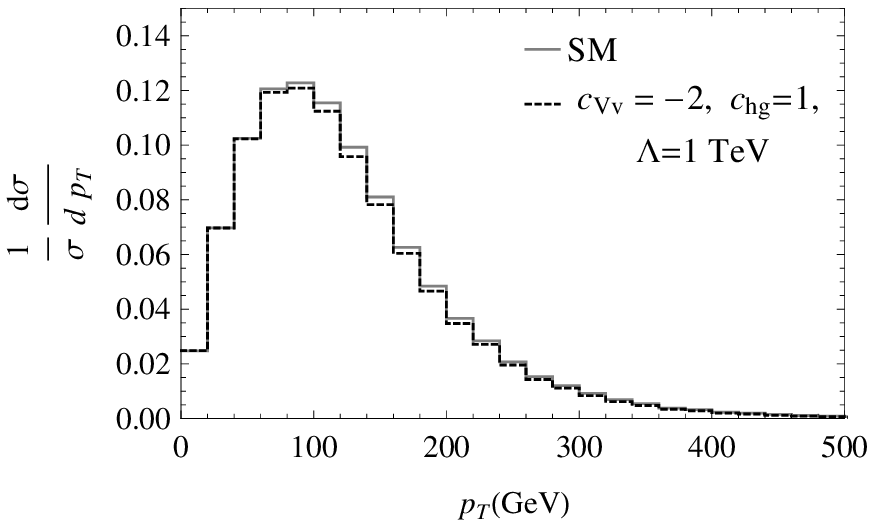}		
	\includegraphics[width=0.48\textwidth]{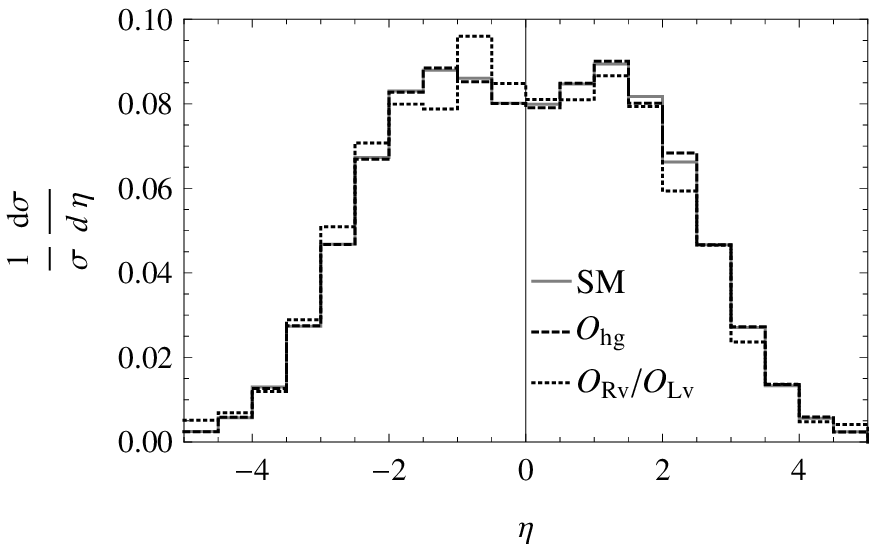}
	\includegraphics[width=0.48\textwidth]{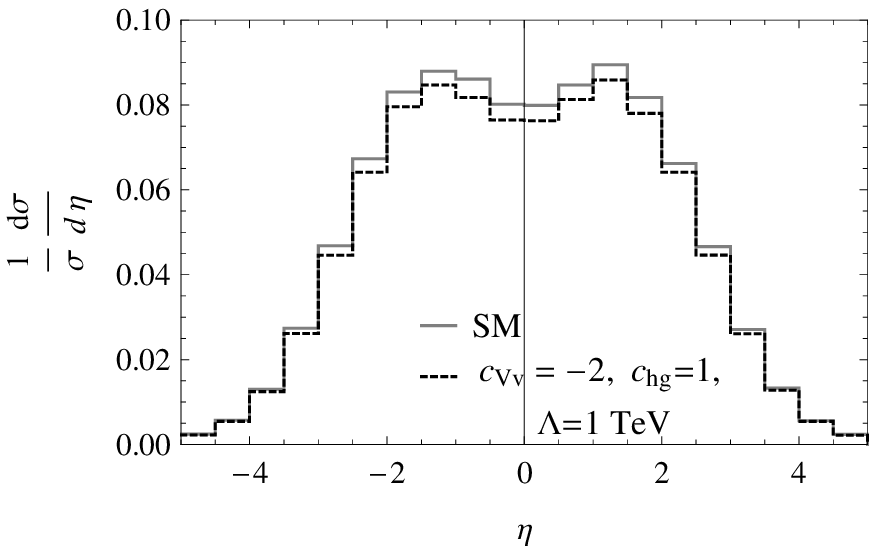}		
	\caption{\small On the left: normalized differential cross~sections of the SM, $\frac{1}{\sigma_{SM}}\frac{d\sigma_{SM}}{dX}$, and of the interferences of the SM with ${\mathcal O}_{hg}$ 
	and with ${\mathcal O}_{Rv}$ and ${\mathcal O}_{Lv}$, $\frac{1}{\sigma_{NP}}\frac{d\sigma_{NP}}{dX}$, as a function of $m_{t\bar{t}}$, $p_T$ and $\eta$ for the LHC at 14TeV. 
	On the right:  normalized cross~section of the SM, $\frac{1}{\sigma_{SM}}\frac{d\sigma_{SM}}{dX}$, and of the SM and the interference with the new physics, $\frac{1}{\sigma_{SM}+\sigma_{NP}}\frac{d\sigma_{SM}+\sigma_{NP}}{dX}$  (for $c_{hg}=1$, $c_{Vv}=-2$ and $\Lambda=1$~TeV).
	}
\label{LHCdistr}
\end{figure*}

The good constraints obtained with the invariant mass at the Tevatron suggest to look for similar effects at the LHC. However, at the LHC, the  top pair is mainly produced by gluon fusion and the contributions of ${\mathcal O}_{Rv}$ and ${\mathcal O}_{Lv}$ are much smaller than the SM contribution. Moreover, the effect of these operators becomes important at high energy where our expansion breaks down. Only ${\mathcal O}_{hg}$ has an important contribution. However, this contribution has a similar shape as that of the SM for reasons already mentioned in Section~\ref{sec:partonic} and confirmed by Fig.~\ref{LHCdistr}. The effects of the new operators will be much harder to be seen in the $m_{t\bar{t}}$ distribution but also in the $p_T$ and $\eta$ at the LHC, as shown in Fig.~\ref{LHCdistr}. 

%
%

\subsection{Forward-backward asymmetry}
\label{sec:AFB}

\begin{figure*}[!bht]
        \centering
                \includegraphics[width=0.48\textwidth]{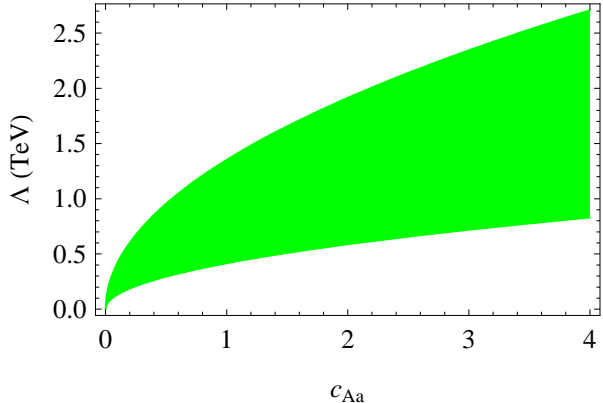}        
                \caption{        \label{Afb}
\small Region of parameter space that can explain the $A_{FB}$ measurement at the Tevatron at one $\sigma$ for $c'_{Aa}=0$.  }
\end{figure*}

In this section we analyse the forward-backward asymmetry in our framework (for an analogous study with older data see Ref.~\cite{Jung:2009pi}). The forward-backward asymmetry in $t \bar t$ production is defined as
\begin{equation}
A_{FB}\equiv \frac{\sigma\left(\cos\theta_t>0\right)-\sigma\left(\cos\theta_t<0\right)}{\sigma\left(\cos\theta_t>0\right)+\sigma 
\left(\cos\theta_t<0\right)}
\end{equation}
where $\theta_t$ is the angle between the momenta of the incoming parton and the outgoing top quark in the laboratory frame.
In the Standard Model, there are no preferred directions for the top and anti-top quarks at the lowest order. A positive asymmetry is generated at NLO, {\it i.e.}, top quarks prefer to go in the direction of the incoming quark and the anti-top quarks in the direction of the incoming antiquark~\cite{Antunano: 
2007da}:
\begin{equation}
A_{FB}^{\rm SM}=0.05\pm0.015.
\end{equation}
The recent measurement of $A_{FB}$ at the Tevatron shows an intriguing deviation from the Standard Model prediction~ \cite{Schwarz:2006ud,:2007qb, Aaltonen:2008hc}.
The most recent CDF result (5.3 fb$^{-1}$)~\cite{CDFNoteAFB}
\begin{equation}
A_{FB}^{\rm EXP}= 0.15\pm0.05(\text{stat}) \pm0.024(\text{syst}),
\end{equation}
is larger by about 2$\sigma$  than the SM prediction.
While a thorough investigation within the Standard Model and in particular of the impact of the unknown higher order QCD corrections would be certainly welcome, it is tempting to explain this discrepancy as the effect of new physics in various models~\cite{Djouadi:2009nb, Jung:2009jz, Cheung:2009ch, Frampton:2009rk, Shu:2009xf, Arhrib:2009hu, Dorsner:2009mq, Jung:2009pi, Cao:2009uz, Barger:2010mw, Cao:2010zb, Chivukula:2010fk, Bauer:2010iq}. An attractive, simple and model-independent alternative is to consider the low energy effective field theory of Section 2. 
A first obvious observation is that no asymmetry can arise in gluon fusion in which the initial state is symmetric.
From Eq.~\eqref{qq}, we see that the asymmetry can only depend on $c_{Aa}$ and $c'_{Aa}$. Since their contribution is a purely odd function of the scattering angle $\theta$ defined in Eq.~\eqref{tcosthetarelation}, these coefficients are only constrained by the asymmetry and not by the total cross~section nor the invariant mass distribution.
After integration with the pdf, we find in the lab frame
\begin{equation}
   \sigma\left(\cos\theta_t>0\right)-\sigma\left(\cos\theta_t<0\right)  
= \left(0.235^{+0.067}_{-0.042}\,  c_{Aa} + 0.088^{+0.024}_{-0.016}\,  c_{Aa}^\prime \right)\times \left(\frac{\text{1 TeV}}{\Lambda}\right)^2 \ \textrm{pb}
\end{equation}
where again the errors are estimated by varying the factorisation and renormalisation scales. Assuming that the total cross~section is given by Eq.~\eqref{xst}, the correction to the SM asymmetry can be expressed  as
\begin{equation}
\delta A_{FB}^{\dim 6}=\left(0.0342^{+0.016}_{-0.009}\, c_{Aa} + 0.0128^{+0.0064}_{-0.0036}\, c_{Aa}^\prime \right) \times \left(\frac{\text{1 TeV}}{\Lambda}\right)^2    \qquad {\rm (Tevatron)}.
\end{equation}
We see once again that the leading contribution comes from the isospin-0 operators. The region of parameter space in the $(c_{Aa}, \Lambda)$ plane that can explain the $A_{FB}$ for $c_{Aa}^\prime=0$ is shown in Fig.~\ref{Afb}. 

It is  instructive to link the simple analysis given above with models featuring an axigluon $A$, {\it  i.e.}, a massive color octet gauge boson coupled to chiral fermionic currents. These models do generate a forward-backward asymmetry due to the interference between the SM amplitude and that of  $q \bar q \to A \to t\bar t$. If the scattering energies are smaller than the mass of the axigluon, the interference terms exactly match the term in Eq.~(\ref{qq}) proportional to $c_{Aa}$. If the axigluon has a flavour-universal coupling to fermions with a strength proportional to the QCD couplings, $g_s$, as in Ref.~\cite{Antunano:2007da}, then the relation $c_{Aa}/\Lambda^2 = -2 g_s^2/m_A^2$ (where $m_A$ is the axigluon mass) obviously leads to a negative asymmetry. To generate a positive asymmetry that could explain the Tevatron result, a flavour non-universal axigluon is needed. More precisely, the coupling of the axigluon to the third generation and to the light quarks should be of opposite sign~\cite{Ferrario:2009bz, Frampton:2009rk, Chivukula:2010fk}: $c_{Aa}/\Lambda^2 = -2 g_A^q g_A^t/m_A^2$ is then positive and can potentially explain the Tevatron data for a mass of the axigluon around 1.5~TeV provided that its couplings are of the same order as the QCD coupling.\footnote{It has been noted~\cite{Chivukula:2010fk} recently that concrete realizations of this axigluon idea~\cite{Frampton:2009rk} are endangered by data on neutral $B_d$-meson mixing.}
We also note that models where a flavour-violating $Z'$  is exchanged
in the t-channel in $q\bar q \to t \bar t$ , have a chance to give a positive
asymmetry only if the $Z'$ is relatively light~\cite{Jung:2009jz}. Indeed, in the heavy regime ($m_{Z'}\gg m_t$), the contribution of the $Z'$ to the top pair production is fully captured in terms of our effective Lagrangian with in particular $c_{Aa}/\Lambda^2=-({g_q^L}^2+{g_q^R}^2)/m_{Z'}^2$, where $g_q^i$ denotes the coupling of $Z'$ to the flavour-off diagonal current $\bar{t}_i \gamma^\mu q_i$. Therefore it leads to a negative asymmetry.

\begin{figure}[!ht]
	\centering
		\includegraphics[width=0.5\textwidth]{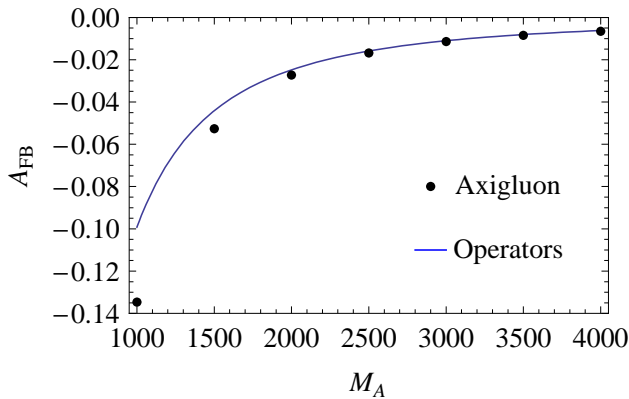}
		\caption{\label{fig:Comparison2}
\small $A_{FB}$ prediction at the Tevatron due to  an axigluon and comparison with the effective field theory approach.
 }
\end{figure}

In Fig.~\ref{fig:Comparison2}, we plot the prediction for $A_{FB}$ from an axigluon with coupling $g_s$ to all fermions and the prediction obtained with  the corresponding effective operator ($C_{Aa}=-2g_s^2$, $C^{\prime}_{Aa}=0$, $\Lambda=M_A$). This shows that our effective field theory approach is a good approximation at the Tevatron for masses $M_A \gtrsim 1.5$ TeV, comparably to the LHC (see Fig.~\ref{fig:Comparison}).

Finally, as an illustration of the simplicity of such an approach, we consider the forward-backward asymmetry at LHC. In this case the symmetry of the $pp$ collision and the dominance of the $gg$ channel for $t \bar t$ make  it particularly challenging. A possibility is to build the so-called central rapidity asymmetry
\begin{equation}
A_{C}(y_C) \equiv \frac{\sigma_t\left(|y|<y_C\right)-\sigma_{\bar t} 
\left(|y|<y_C\right)}{\sigma_t\left(|y|<y_C\right)+\sigma_{\bar t}\left(|y|<y_C\right)}  \qquad {\rm (lab \, frame)}\,,
\end{equation}
where $y_C$ is the rapidity cut defining the ``centrality'' of an event. The value $y_C=1$ has been shown to be close
to optimal in Ref.~\cite{Antunano:2007da}. A straightforward calculation using  $c_{Aa}\left(\frac{\text{1 TeV}} 
{\Lambda}\right)^2=2$ as a central extraction from the  Tevatron data gives rise to very small asymmetries, $A_{C}\lesssim 1 \%$, at the LHC both at 14 TeV and 7 TeV. While the effects of new physics could be enhanced by requiring, for instance, a minimal invariant $t\bar t$ mass, it is also clear that measurements of forward-backward asymmetries will be very challenging at the LHC.

\subsection{Spin correlations}
\label{sec:spins}

We are here focussing on spin correlations which can provide further information on the coupling structure of the production mechanism (for alternative approaches see Ref.\cite{Godbole:2010kr}). Spin correlations are good observables to disentangle the contributions from the two operators ${\mathcal O}_{Rv}$ and ${\mathcal O}_{Lv}$ since at high energy ${\mathcal O}_{R/Lv}$ should produce mainly right/left-handed tops and left/right-handed antitops. 

\begin{figure*}[htb!]
	\centering
	\includegraphics[width=0.48\textwidth]{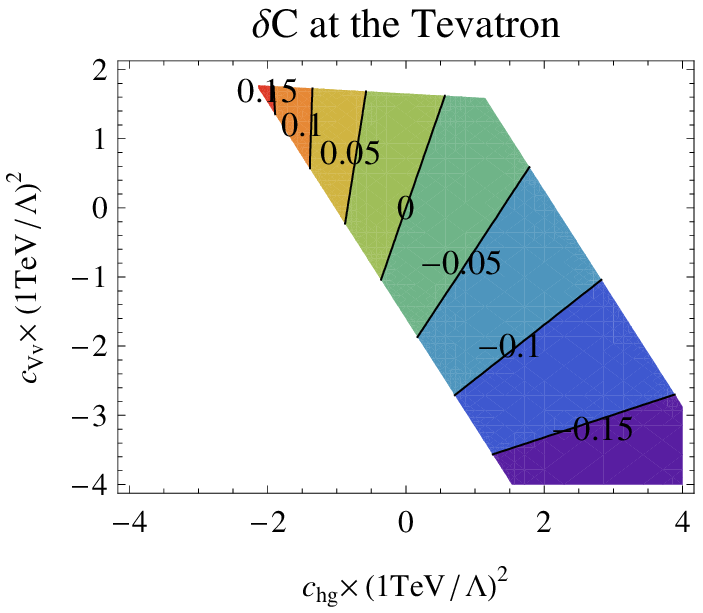} 
		\includegraphics[width=0.48\textwidth]{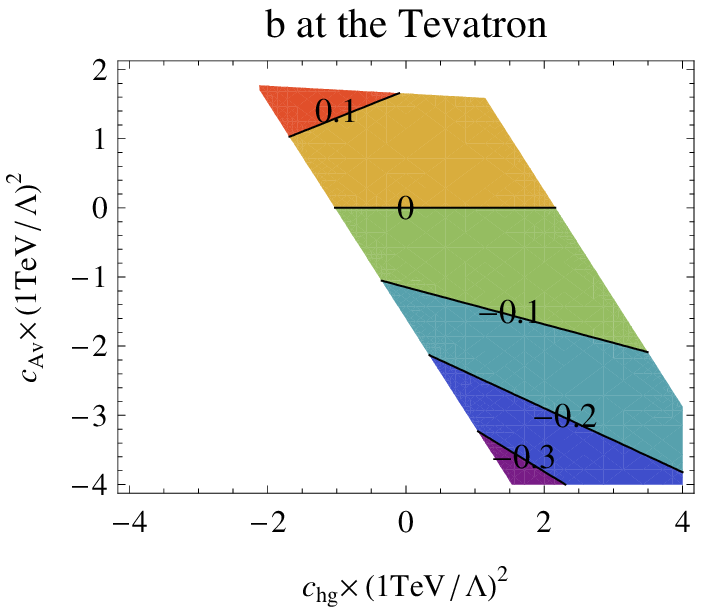}
		\includegraphics[width=0.45\textwidth]{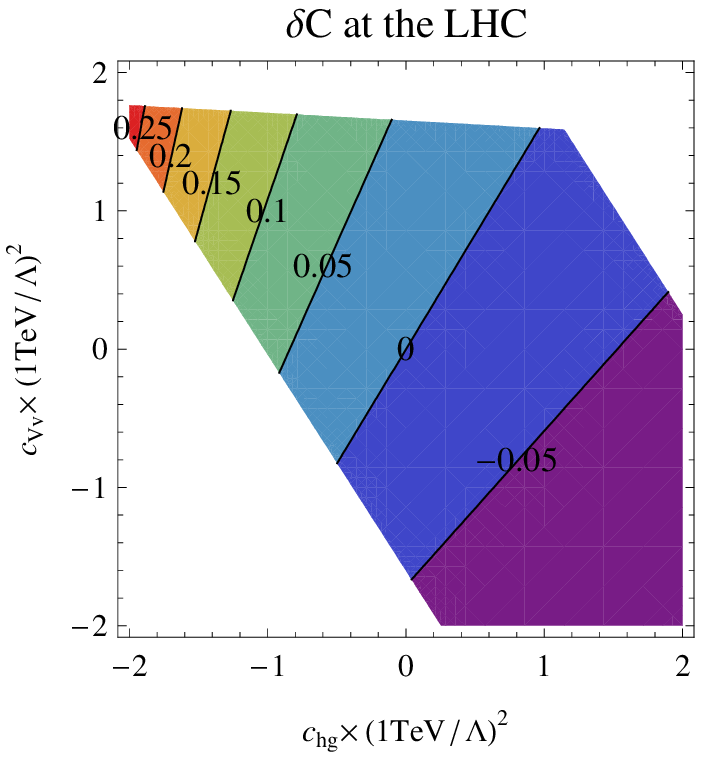}
		\hspace{.5cm}
		\includegraphics[width=0.45\textwidth]{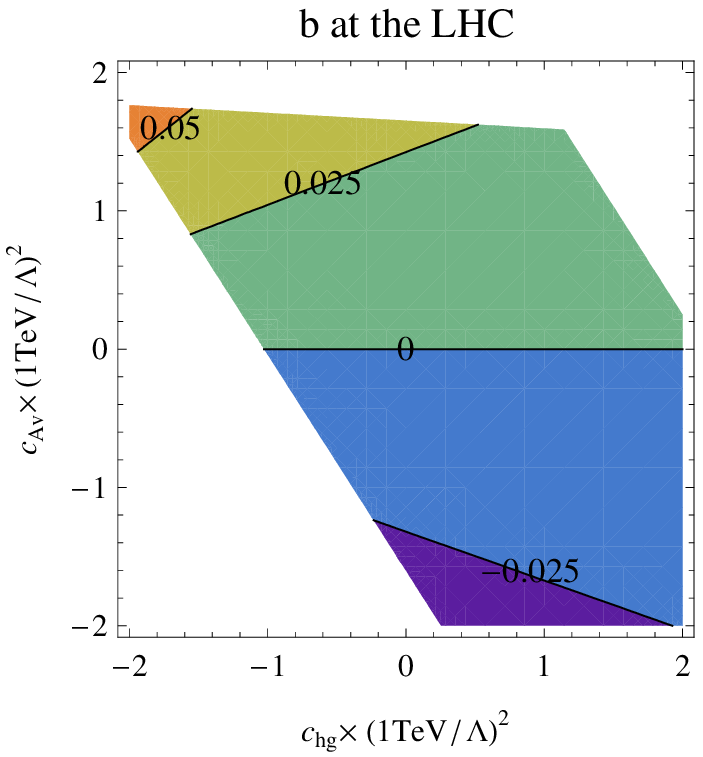}		
		\caption{	\label{spintev}
		\small {\it Top panel:} Deviations  from the SM prediction
		at the Tevatron ($C=0.47$, $b=0$)  \cite{Bernreuther:2004jv} for the parameters $C$ (on the left) and $b=b_+=b_-$ (on the right) in the region allowed by the Tevatron.  {\it Bottom panel:} Deviations at the LHC from the SM prediction ($C=-0.31$, $b=0$) \cite{Bernreuther:2004jv}.}
\end{figure*}

In fact, there is only one dimension-six operator affecting the top decay, $\left(H\bar{Q}\right)\sigma^{\mu\nu}\sigma^ItW_{\mu\nu}^I$, which however does not modify the maximal spin-correlation in the leptonic decays of the top quark~\cite{Zhang:2010dr,Grzadkowski:1999iq,Grzadkowski:2002gt}, {\it i.e.}, 
\begin{equation}
\frac{\Gamma_\uparrow}{\Gamma}=\frac{1+\cos \theta}{2},\qquad \frac{\Gamma_\downarrow}{\Gamma}=\frac{1-\cos \theta}{2},
\end{equation}
where $\theta$ is the angle between the charged lepton and the spin of the top quark and the arrows denote the different projections of the top spin. Consequently, the general form of the normalized differential $t\bar{t}$ cross~section is given by
\begin{equation}
\frac{1}{\sigma}\frac{d\sigma}{d\cos\theta_+d\cos\theta_-} = \frac{1}{4}\left(1+C \cos\theta_+ \cos\theta_- +b_+\cos\theta_+ + b_-\cos\theta_-\right),
\end{equation}
where $\theta_+$ $(\theta_-)$ is the angle between the charged lepton $l^+$ ($l^-$)  resulting from the top (antitop) decay and some reference direction $\vec{a}$ ($\vec{b}$). For this study, we chose the helicity basis, $\vec{a}=-\vec{b}=\vec{p_1}$ where $\vec{p_1}$ is the top momentum in the $t\bar{t}$ rest frame\footnote{It has been shown~\cite{Mahlon:1995zn} that spin correlation effects in the SM are more important at the Tevatron in the beam basis. However, it appears that the deviations from the SM values due to the operators ${\mathcal O}_{hg}$, ${\mathcal O}_{Rv}$ and ${\mathcal O}_{Lv}$ are on the contrary smaller in the beam basis.}. There is a one-to-one relation between the parameters $C$ and $b_\pm$ and the helicity cross~sections, 
\begin{eqnarray}
C &=& \frac{1}{\sigma}\left(\sigma_{RL}+\sigma_{LR}-\sigma_{RR}-\sigma_{LL}\right),\\
b_+ &=& \frac{1}{\sigma}\left(\sigma_{RL}-\sigma_{LR}+\sigma_{RR}-\sigma_{LL}\right),\\
b_- &=& \frac{1}{\sigma}\left(\sigma_{RL}-\sigma_{LR}-\sigma_{RR}+\sigma_{LL}\right).
\end{eqnarray}
The explicit formulas for the helicity cross~sections are given in App.~\ref{app:xsec} and lead to (neglecting the contributions from the isospin-1 sector):\\
\begin{eqnarray}
C \times  \sigma /\mbox{pb} &\!\!\!  = \!\!\! & 2.82^{+1.06}_{-0.72}+\left[\left(0.37^{+0.10}_{-0.08}\right) c_{hg}+\left(0.50^{+0.13}_{-0.10}\right) c_{Vv} \right] \times \left(\frac{1\mbox{ TeV}}{\Lambda} \right)^2\!\!,\\
b \times \sigma  /\mbox{pb} & \!\!\! = \!\!\!  & \left(0.45^{+0.12}_{-0.09}\right)  c_{Av} \times \left(\frac{1\mbox{ TeV}}{\Lambda} \right)^2\!\!,
\end{eqnarray}
at the Tevatron, and
\begin{eqnarray}
C \times  \sigma /\mbox{pb} &\!\!\!  = \!\!\! &  -166^{+52}_{-37}+ \left[\left(-69^{+17}_{-13}\right) c_{hg}+\left(11^{+1}_{-1}\right) c_{Vv} \right] \times \left(\frac{1\mbox{ TeV}}{\Lambda} \right)^2\!\!,\\
b \times \sigma  /\mbox{pb} & \!\!\! = \!\!\!  &\left( 10^{+1}_{-1}\right)   c_{Av} \times \left(\frac{1\mbox{ TeV}}{\Lambda} \right)^2\!\!,
\end{eqnarray}
at the LHC. The parameters $b_\pm$ are exactly proportional to the difference $c_{Rv}-c_{Lv}$ and thus allow us to distinguish between right or left handed top quarks. Additionally, the parameter $C$ depends quite strongly on $c_{hg}$ and $c_{Vv}$ and can be used to detect the presence of new physics as shown in Fig.~\ref{spintev} for the Tevatron and the LHC respectively. The errors on the contour lines are only of a few percents.

%

\begin{figure*}[htb!]
	\centering
		\includegraphics[width=0.45\textwidth]{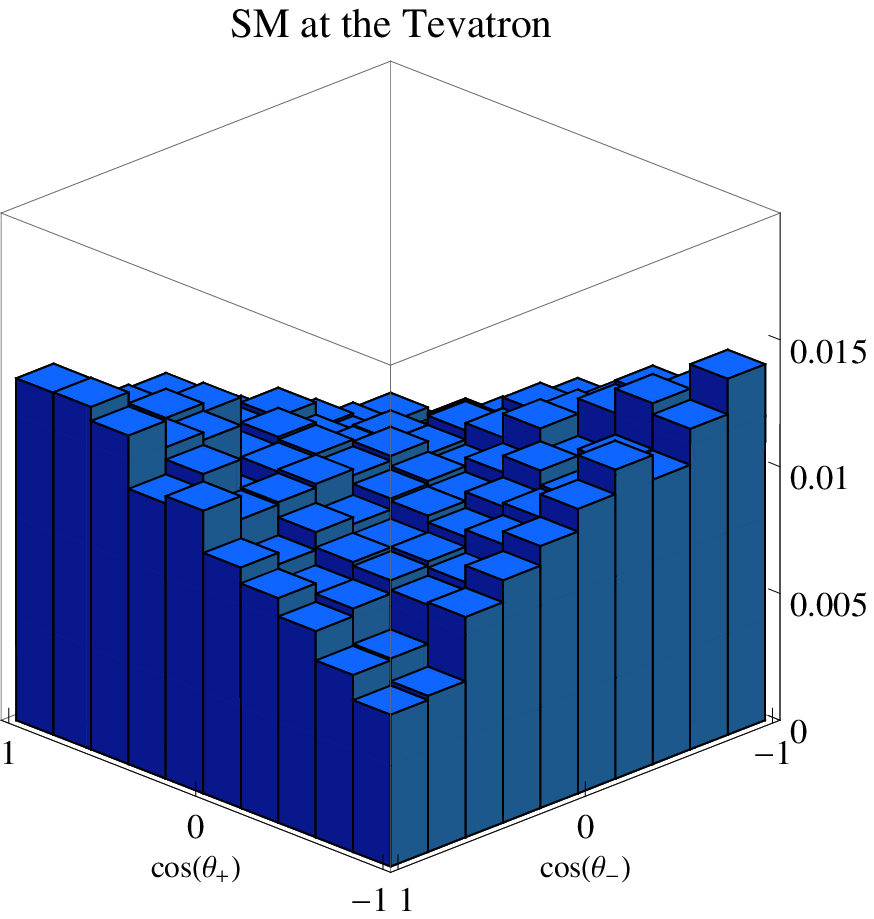}
		\hspace{.5cm}
		\includegraphics[width=0.45\textwidth]{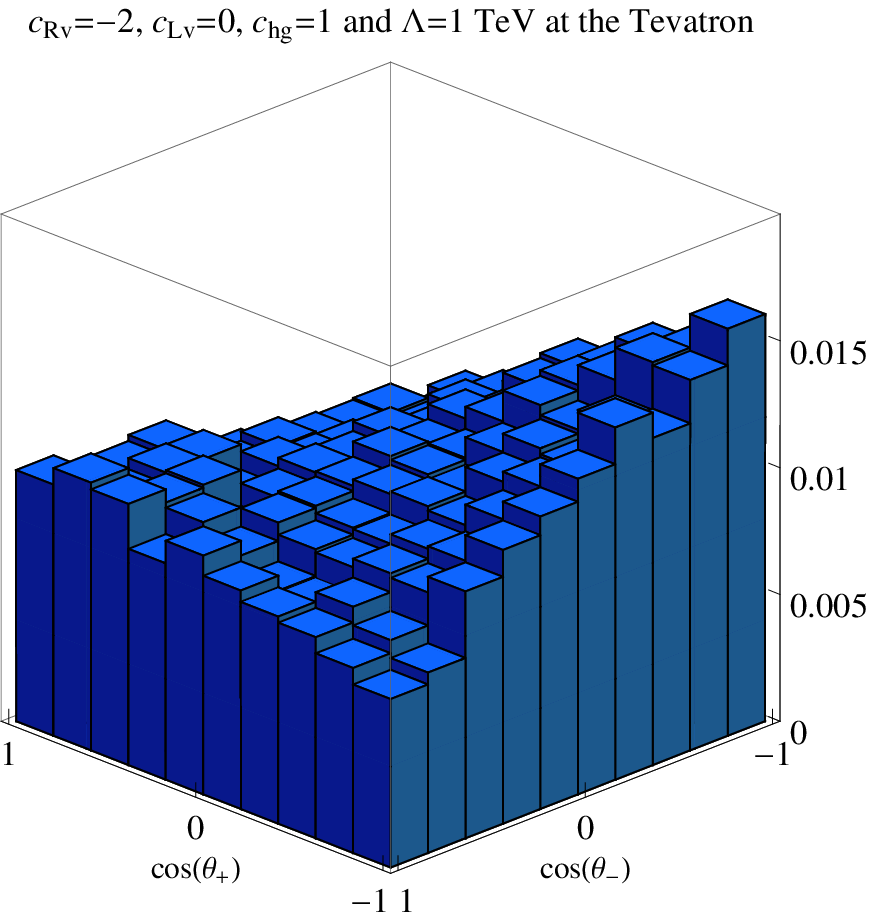}	
		\includegraphics[width=0.45\textwidth]{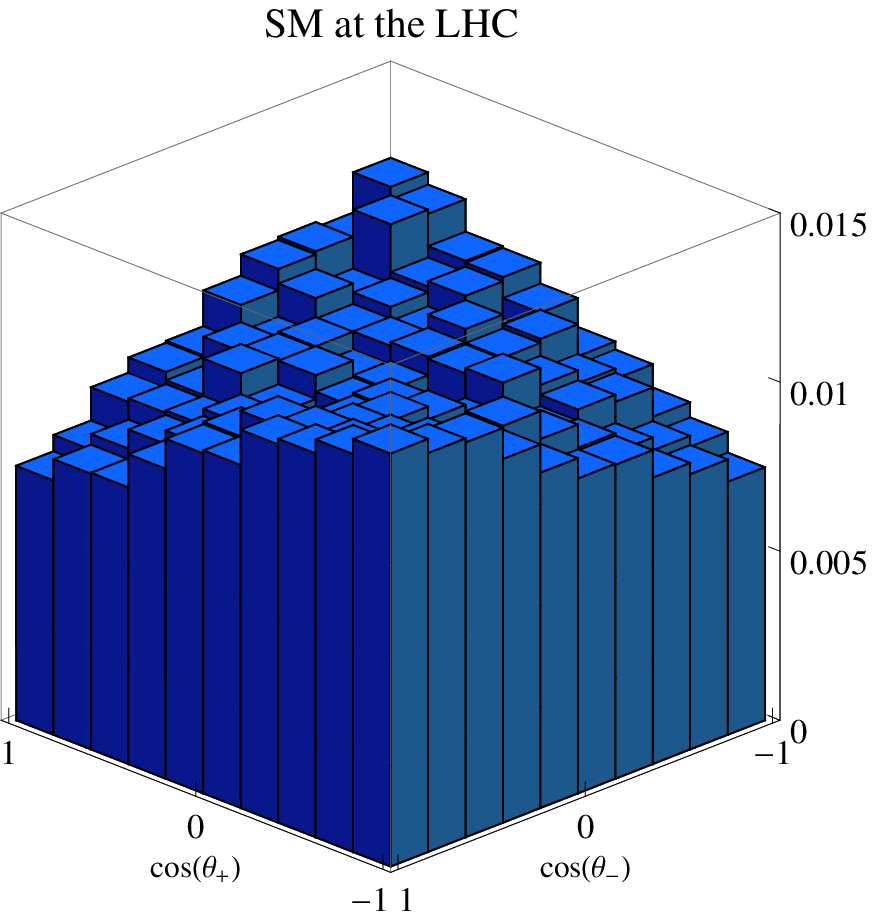}
		\hspace{.5cm}
		\includegraphics[width=0.45\textwidth]{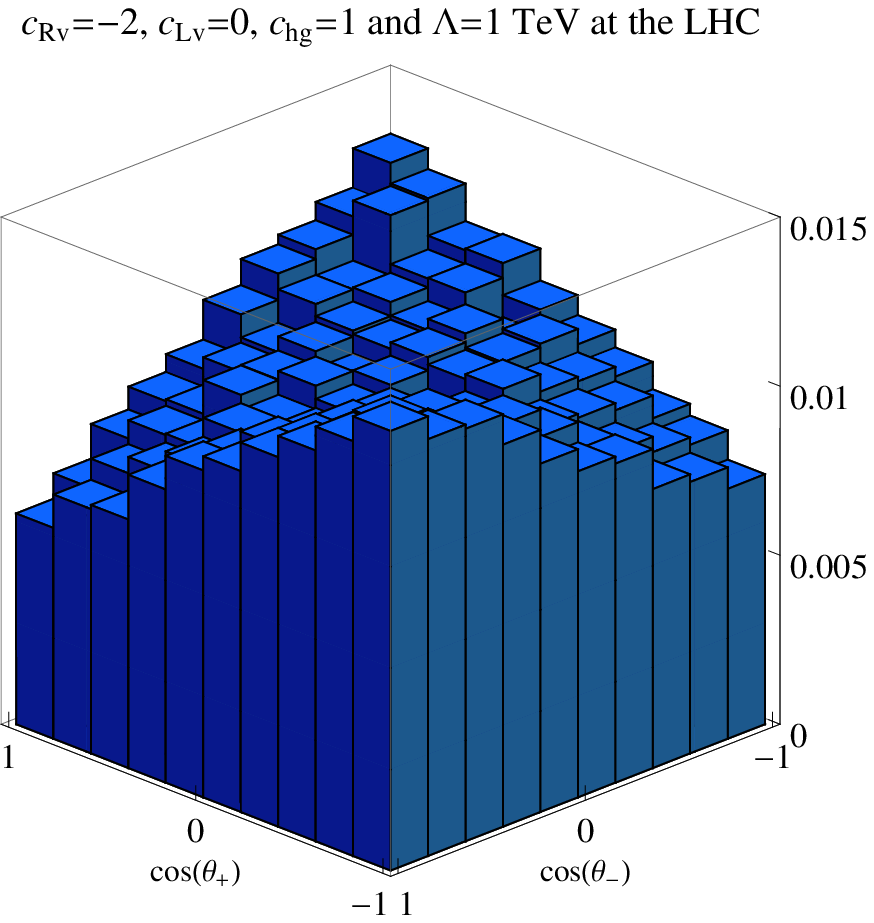}		
		\caption{	\label{spindist}
		\small Distribution of events at the Tevatron/LHC (top panel/bottom panel) for the SM (on the left) and for $c_{Rv}=-2$, $c_{Lv}=0$, $c_{hg}=1$ and $\Lambda=1$~TeV (on the right) with $\mu_F=\mu_R=mt$. }
\end{figure*}

As expected, the parameters $b=b_+=b_-$ only differ slightly from zero at the LHC where the contributions of $\mathcal{O}_{Rv}$ and $\mathcal{O}_{Lv}$ are small. A possible modification of the spin distribution both at the Tevatron and the LHC is shown in Figs.~\ref{spindist}. However, it will be quite difficult to measure them at the Tevatron where only a few hundreds of events are expected and observed (Ref.~\cite{Jabeen:2009nu} and Ref.~[7] therein),  while at the LHC we expect about a few millions of events after 100~fb$^{-1}$~\cite{Bayatian:2006zz, :1999fr}.  

%


\section{$t\bar{t}b\bar{b}$ and $t\bar{t}t\bar{t}$ production at the LHC}
\label{sec:4t}

While $t\bar{t}$ production appears as the leading process to probe any new physics in the top sector, there are physical situations where the operators of $\mathcal{L}_{t\bar{t}}$ are parametrically suppressed. As shown in Section~\ref{sec:Composite}, this is the case if the top quark is not an elementary particle but rather a composite bound state: the dominant operators are the ones involving composite states only. The $t\bar{t}$ process is still probing the dominant operators but at the loop level only. In these situations, a much better probe of the dominant dynamics (Eqs.\eqref{odr} to~\eqref{odb})  is the direct production of four top-quarks or the production of two top- and two bottom-quarks~\cite{Pomarol:2008bh}.

\begin{figure}[h!]
	\centering
		\includegraphics[width=0.64\textwidth]{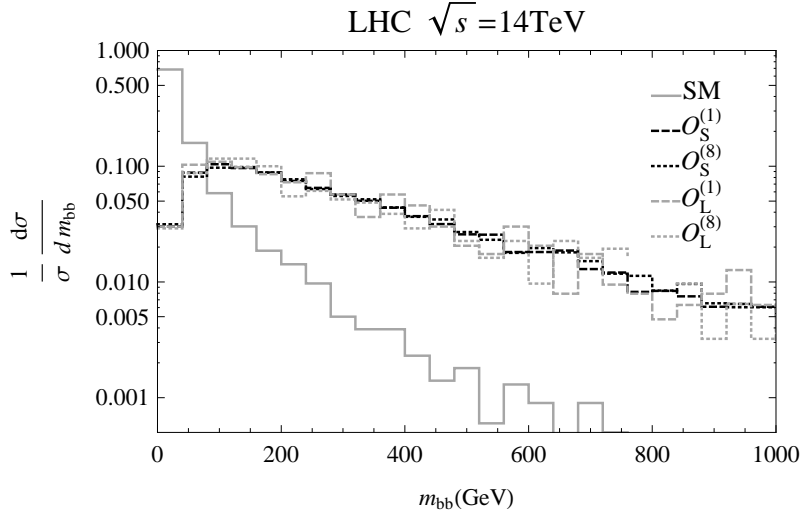}
		\caption{\small Normalized cross~sections at the LHC for the $t\bar{t}b\bar{b}$ production as a function of the $b\bar{b}$ invariant-mass.  The continuous line represents the SM while the dark dashed/dotted lines represent the contributions of the scalar color-singlet operator $\mathcal{O}_S$/color-octet operator $\mathcal{O}^8_S$ and the gray dashed/dotted lines represent the contributions of the vector color-singlet operator  $\mathcal{O}_V$/color-octet operator  $\mathcal{O}^8_V$. Since we neglected the interference terms between the SM and the new physics contribution, the distributions are independent of the new physics scale and of the actual couplings in front of the dimension-6 operators.}
	\label{minv2b2t}
\end{figure}

\begin{figure}[h!]
	\centering
		\includegraphics[width=0.64\textwidth]{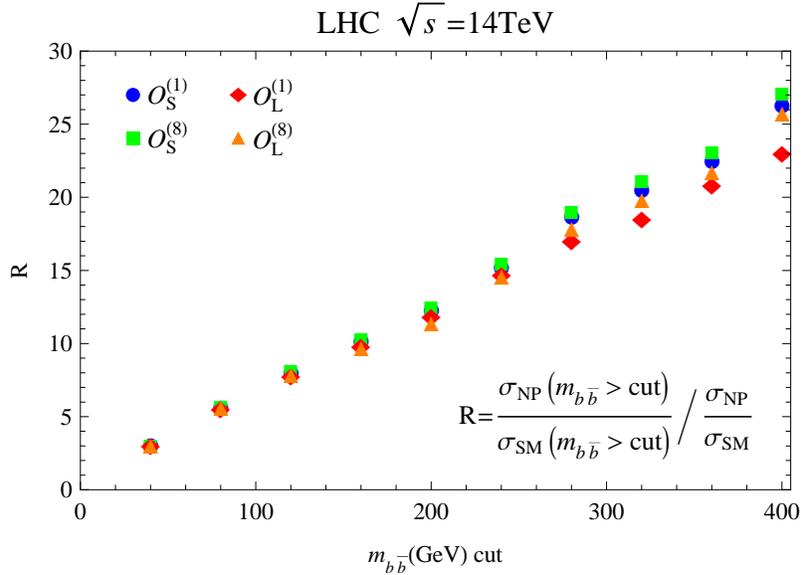}
		\caption{\small Effect of the $b\bar{b}$ invariant-mass cut on the signal over background ratio. $R=\frac{\sigma_{NP}(m_{b\bar{b}}>cut)}{\sigma_{SM}(m_{b\bar{b}}>cut)}/\frac{\sigma_{NP}}{\sigma_{SM}}$ is the double ratio of the signal (contribution from new operators) over the background (contribution of the SM) with and without  the cut on the $b\bar{b}$ invariant mass. In our approximations, $R$ is independent of the new physics scale and of the actual couplings in front of the dimension-6 operators.}
	\label{sob}
\end{figure}

The SM cross~section for 4-top production is rather small (of the order of 5~fb at the LHC) and the operators of Eqs.~(\ref{odr})--(\ref{odb}) can easily give larger contributions. Contrary to pair production, the smallness of the SM cross~section urges us to keep the squared amplitude from new physics instead of the interference with the SM  as shown in Table~\ref{tab:ratio}. The two contributions are equal for $0.25 <c_i\left(\frac{\textrm{1 TeV}}{\Lambda}\right)^2<0.75$ where $c_i$ generically denotes the coefficient of the operator $\mathcal{O}_i$. The range of this critical value is due to the different operators. Thus, we are effectively computing the cross~sections at the order $\mathscr{O}(\Lambda^{-4})$ and we also neglect the interference between the SM and any dimension-8 operators. 
\begin{table}[!h]
\centering
 \begin{tabular}{|l|r|r|r|r|r|r|r|r|}
 \hline
  &$\sigma_{4t}$&$\sigma_{4t}^{\Lambda^{-2}}$&$\sigma_{4t}^{\Lambda^{-4}}$ &$\sigma_{t\bar{t}b\bar{b}}$&$\sigma_{t\bar{t}b\bar{b}}^{\Lambda^{-2}}$&$\sigma_{t\bar{t}b\bar{b}}^{\Lambda^{-4}}$&$\sigma_{t\bar{t}b\bar{b}}^{\textrm{cut}}$&$\sigma_{t\bar{t}b\bar{b}}^{\textrm{cut}}/\sigma_{4t}$\\[0.1cm]
 	 &  (fb) & (fb) & (fb) & (pb) & (pb) & (pb) & (pb) & \\
	\hline
 	SM & $4.86$ & - & - & 7.2 & - & - & 0.348 & 71.6\\
 	${\mathcal O}^{(1)}_{R}$ & - & 2.7 & 138 & - & - & - & - & -\\ 
 	${\mathcal O}^{(1)}_{S}$ & - & 2.9 & 48 & - &$<$1.1& 7.60 & 4.40 & 92\\ 
 	${\mathcal O}^{(8)}_{S}$ & - & 0.49 & 11 & - &$<$0.2& 1.28 & 0.76 & 71\\ 
 	${\mathcal O}^{(1)}_{L}$ & - & 2.7 & 138 & - &$<$0.5& 3.61 & 2.12 & 15.6\\ 
 	${\mathcal O}^{(8)}_{L}$ & - & 0.91 & 15 & - &0.49& 0.77 & 0.42 & 28.2\\ 
	\hline
	\end{tabular}
\caption{\small The $t\bar{t}t\bar{t}$ ($\mu_F=\mu_R=4m_t$) and $t\bar{t}b\bar{b}$ ($\mu_F=\mu_R=2m_t$) cross-sections for $\Lambda=1$~TeV and $c_i=4\pi$. The interferences between the SM and the new physics, given in the third and sixth column, can be neglected. The squared amplitudes from new physics are in the fourth and seventh column. The new physics contributions for a different scale $\Lambda$ and different couplings $c_i$ are simply obtained by multiplying the numbers given above by a factor $(c_i/(4\pi))^2\times (\textrm{1 TeV}/\Lambda)^4$. }
	\label{tab:ratio}
\end{table} 
The SM $t\bar{t}b\bar{b}$ production is not as suppressed as the 4-top production, so the same approximation would be a priori valid only for smaller values of the scale of new physics (or for larger couplings). However, we can make use of the particular kinematics associated to the new physics operators to improve our approximation. Indeed, the new physics squared amplitudes grow with the energy as shown in Fig.~\ref{minv2b2t}. Therefore the $b\bar{b}$ pair will be produced with a higher invariant-mass in presence of new physics, and a cut on the $b\bar{b}$ invariant-mass will suppress the $\mathscr{O}(\Lambda^{-2})$ terms compared to the 
$\mathscr{O}(\Lambda^{-4})$ ones. Moreover, such a cut will also improve the ratio of the signal over the SM background as shown on Fig.~\ref{sob}.

For both $t\bar{t}t\bar{t}$ and $t\bar{t}b\bar{b}$ production, the operators defined in Eqs.~(\ref{odr})--(\ref{odb}) give cross~sections of the same order of magnitude (see Table~\ref{tab:ratio}) and it is not possible to distinguish them just by a measurement of the total cross-section. Furthermore, as Fig.~\ref{minv2b2t} suggests, they also generate similar distributions for all the spin-independent variables. However, the ratio of the two cross~sections appears to be very different for the different operators and it is also independent of the new physics scale and of the actual couplings in front of the dimension-6 operators provided that the interferences with the SM contribution can be safely neglected.  A detailed study of four-top production at the LHC will be presented in Ref.~\cite{gauthierServant} (see Ref.~\cite{Brooijmans:2010tn} for a preview).

\section{Summary}
\label{sec:conclusion}

In theories that provide a mechanism for mass generation, new physics must have a large coupling to the top quark. It is therefore natural to use top quark observables to test the mechanism responsible for electroweak symmetry breaking.
We have shown how non-resonant top-philic new physics can be probed using measurements in top quark pair production at hadron colliders. 

Some of our results already appeared in the literature, although only subsets of dimension-six  operators were considered. For instance, there is an extensive literature~\cite{Atwood:1994vm,Cheung:1995nt,Whisnant:1997qu,Hikasa:1998wx,Hioki:2009hm,Choudhury:2009wd} on the operator $\mathcal{O}_{hg}$, the chromomagnetic dipole moment of the top quark,  while  other works focused on the effect of additional four-fermion operators on top pair production at the Tevatron~\cite{Lillie:2007hd,Kumar:2009vs,Jung:2009pi,Cao:2010zb}. Recently, all relevant operators were properly accounted for in Ref.~\cite{Zhang:2010dr} which however did not cover the corresponding phenomenological analysis. In our work,  the aim is to provide a complete and self-consistent treatment  in a model-independent approach and, especially, to extract the physics by combining information from the Tevatron and the LHC.

The analysis can be performed in terms of eight operators, suppressed by the square of the new physics energy scale $\Lambda$. 
Observables  depend on different combinations of only four main parameters 
\begin{center}
\begin{tabular}{lll}
$\sigma (gg\rightarrow t\bar{t}) ,{d\sigma (gg\rightarrow t\bar{t})}/{dt} $& $\leftrightarrow$&$c_{hg}$\\
&&\\
$\sigma (q\bar{q}\rightarrow t\bar{t})$& $\leftrightarrow$&$c_{hg},c_{Vv}$\\
&&\\
${d\sigma (q\bar{q}\rightarrow t\bar{t})}/{dm_{tt}} $& $\leftrightarrow$& $c_{hg},c_{Vv}$\\
&&\\
$A_{FB}$& $\leftrightarrow$& $c_{Aa}$\\
&&\\
$\mbox{spin correlations}$ & $\leftrightarrow$& $c_{hg},c_{Vv},c_{Av}$\\
\end{tabular}
\end{center}
where $c_{hg}$ is the parameter associated with the chromomagnetic dipole moment operator and $c_{Vv}$, $c_{Aa}$, $c_{Av}$ correspond to particular combinations of four-fermion operators defined in Section \ref{sec:relop}. Let us summarize our main results on these observables.
\begin{enumerate}
 \item Since top pairs are mainly produced by gluon fusion at the LHC, the measurement of the $t\bar{t}$ cross-section at the LHC will determine the allowed range for $c_{hg}$. In contrast, the Tevatron cross~section is also sensitive to the four-fermion operators and constrains a combination of $c_{hg}$ and $c_{Vv}$. Consequently,  the measurements of the total cross~section at the Tevatron and at the LHC are complementary and combining the two  will pin down the allowed region in the $(c_{hg}, c_{Vv})$ plane. We emphasize that the $\mathcal{O}_{hg}$ operator can only be generated at the loop-level in resonance models. Consequently, $c_{hg}$ is expected to be small in such models.

 \item The shape of the invariant mass distribution at the Tevatron is sensitive to a combination of the parameters $c_{Vv}$ and $c_{hg}$ which is different from the combination controlling the total cross section.
It depends quite strongly on the presence of four-fermion operators and was used to further reduce the parameter space mainly along the $c_{Vv}$ direction.

 \item The forward-backward asymmetry that probes different operators than those affecting the cross~section or the invariant mass distribution could be the first sign of new physics at the Tevatron. The scale of the new interaction(s) can then be estimated from the value predicted by our effective Lagrangian approach if a deviation from the SM is confirmed. 
 
 \item The three observables $\sigma$, $d\sigma/dm_{t\bar{t}}$ and $A_{FB}$ are unable to disentangle between theories coupled mainly to right- or left-handed top quarks.
However, spin correlations allow us to determine which chiralities of the top quark couple to new physics, and in the case of composite models, whether one or two chiralities of the top quark are composite.
\end{enumerate}

In composite models, the ratio of $c_{Vv}$ and $c_{hg}$ is very important  since it reflects the number of composite fields in the SM. However, the peculiar hierarchy between dominant and subdominant operators cannot be tested in $t\bar{t}$ production that depends on one class of operators only. Fortunately, composite models can be further tested through the golden four-top channel and $t\bar{t}b\bar{b}$ production at the LHC. Both processes are necessary to identify the dominant operators and thus to extract their coefficients. The hierarchy between the operators can be tested and used to estimate the strength of the new strong interaction,~$g_\rho$. 
We stress that the results for top pair production are generic while those for $t\bar{t}t\bar{t}$ and $t\bar{t}b\bar{b}$ production require the enhancement due to a strong interaction. These two processes would disappear in the SM background if they are not enhanced by a factor $g_\rho^2$.

Finally, we stress that in the most recent compositeness scenarios, other mechanisms could lead to $t\bar{t} + X$ final states, such as the decays of fermionic top partners, adding to the richness and interest of these final states. Studying higher dimensional operators capturing these effects could be interesting. We leave this for future investigation.

\section*{Acknowledgments}

This work was started during the 2009 CERN Theory Institute ``Top quark physics: from the Tevatron to the LHC". We thank Tim~Tait for discussions. We also thank Matteo~Cacciari for his help and we are particularly grateful to Nathan~Goldschmidt for providing us the latest CDF data used in~\cite{Goldschmidt:2010}. The work of G.S. as well as C.D.'s visits to CERN were supported by  the European Research Council Starting Grant Cosmo@LHC. C.G. has been partly supported by the European Commission under the contract ERC advanced grant 226371 MassTeV and the contract PITN-GA-2009-237920, UNILHC. The work of C.D., {J.-M.}~G. and F.M. was supported by the Belgian Federal Office for Scientific, Technical and Cultural Affairs through the Interuniversity Attraction Pole No. P6/11. C.D. is a fellow of the Fonds National de la Recherche Scientifique.

\appendix
\section{Fierz transformations}
\label{app:Fierz}

We are collecting here some Fierz transformations that are needed to reduce the basis of independent dimension-six operators.
The same transformations are also useful to compute the effective Lagrangian obtained after integrating out some heavy resonances.
\begin{eqnarray}
&&\delta_{ij} \delta_{kl} = \frac{1}{2} \sigma^I_{il} \sigma^I_{kj}  + \frac{1}{2} \delta_{il} \delta_{kj}\, ,\\
&&\delta_{ab} \delta_{cd} = 2 T^A_{ad}T^A_{cb}  + \frac{1}{3} \delta_{ad} \delta_{cb}\, ,\\
&& (\gamma_\mu P_{L/R})_\alpha{}^\beta (\gamma^\mu P_{L/R})_\gamma{}^\delta =  - (\gamma_\mu P_{L/R})_\alpha{}^\delta (\gamma^\mu P_{L/R})_\gamma{}^\beta\, \\[.2cm]
&& (\gamma_\mu P_{R})_\alpha{}^\beta (\gamma^\mu P_{L})_\gamma{}^\delta =  2\, (P_{L})_\alpha{}^\delta (P_{R})_\gamma{}^\beta\, ,\\
&& (P_{L/R})_\alpha{}^\beta (P_{L/R})_\gamma{}^\delta = -\frac{1}{2}\, (P_{L/R})_\alpha{}^\delta (P_{L/R})_\gamma{}^\beta
+\frac{1}{8}\, (\gamma^{\mu\nu}P_{L/R})_\alpha{}^\delta (\gamma_{\mu\nu}P_{L/R})_\gamma{}^\beta
\, ,
\end{eqnarray}
where $P_{L/R}=(1\mp \gamma^5)/2$ are the usual chirality projectors and $\gamma^{\mu\nu}=\frac{1}{2}\, [\gamma^\mu,\gamma^\nu]$.


\section{Feynman diagrams for $t\bar{t}$ production at order $\mathscr{O}\left(\Lambda^{-2}\right)$}
\label{app:Feynman}

At the $\mathscr{O}(\Lambda^{-2})$ order, the two parton-level cross~sections for $t\bar{t}$ production follow from the Feynman diagrams depicted in Fig.~\ref{Fig:gg} and~\ref{Fig:qq}.

\SetScale{0.45}\setlength{\unitlength}{0.45pt}
\begin{figure}[!ht]
\begin{center}
\fcolorbox{white}{white}{
  \begin{picture}(622,600) (99,-91)
    \SetWidth{1.0}
    \SetColor{Black}
    \Gluon(112,472)(160,408){7.5}{6}
    \Gluon(160,408)(112,344){7.5}{6}
    \Gluon(160,408)(240,408){7.5}{6}
    \Line[arrow,arrowpos=0.5,arrowlength=5,arrowwidth=2,arrowinset=0.2](240,408)(288,472)
    \Line[arrow,arrowpos=0.5,arrowlength=5,arrowwidth=2,arrowinset=0.2](288,344)(240,408)
    \Gluon(336,456)(416,456){7.5}{6}
    \Gluon(336,360)(416,360){7.5}{6}
    \Line[arrow,arrowpos=0.5,arrowlength=5,arrowwidth=2,arrowinset=0.2](496,360)(416,360)
    \Line[arrow,arrowpos=0.5,arrowlength=5,arrowwidth=2,arrowinset=0.2](416,360)(416,456)
    \Line[arrow,arrowpos=0.5,arrowlength=5,arrowwidth=2,arrowinset=0.2](416,456)(496,456)
    \Gluon(560,456)(640,456){7.5}{6}
    \Gluon(560,360)(640,360){7.5}{6}
    \Line[arrow,arrowpos=0.5,arrowlength=5,arrowwidth=2,arrowinset=0.2](640,360)(720,360)
    \Line[arrow,arrowpos=0.5,arrowlength=5,arrowwidth=2,arrowinset=0.2](720,456)(640,456)
    \Line[arrow,arrowpos=0.5,arrowlength=5,arrowwidth=2,arrowinset=0.2](640,456)(640,360)
    \Text(304,408)[lb]{\normalsize{\Black{$+$}}}
    \Text(528,408)[lb]{\normalsize{\Black{$+$}}}
    \Gluon(160,184)(112,120){7.5}{6}
    \Gluon(112,248)(160,184){7.5}{6}
    \Gluon(160,184)(240,184){7.5}{6}
    \Line[arrow,arrowpos=0.5,arrowlength=5,arrowwidth=2,arrowinset=0.2](240,184)(288,248)
    \Line[arrow,arrowpos=0.5,arrowlength=5,arrowwidth=2,arrowinset=0.2](240,-88)(192,-24)
    \Gluon(320,136)(400,136){7.5}{6}
    \Gluon(320,232)(400,232){7.5}{6}
    \Line[arrow,arrowpos=0.5,arrowlength=5,arrowwidth=2,arrowinset=0.2](400,232)(480,232)
    \Line[arrow,arrowpos=0.5,arrowlength=5,arrowwidth=2,arrowinset=0.2](480,136)(400,136)
    \Line[arrow,arrowpos=0.5,arrowlength=5,arrowwidth=2,arrowinset=0.2](400,136)(400,232)
    \Gluon(544,232)(624,232){7.5}{6}
    \Gluon(544,136)(624,136){7.5}{6}
    \Line[arrow,arrowpos=0.5,arrowlength=5,arrowwidth=2,arrowinset=0.2](624,232)(624,136)
    \Line[arrow,arrowpos=0.5,arrowlength=5,arrowwidth=2,arrowinset=0.2](704,232)(624,232)
    \Line[arrow,arrowpos=0.5,arrowlength=5,arrowwidth=2,arrowinset=0.2](624,136)(704,136)
    \Vertex(240,184){5.657}
    \Vertex(400,232){5.657}
    \Vertex(624,136){5.657}
    \Gluon(192,-24)(144,-88){7.5}{6}
    \Gluon(144,40)(192,-24){7.5}{6}
    \Line[arrow,arrowpos=0.5,arrowlength=5,arrowwidth=2,arrowinset=0.2](192,-24)(240,40)
    \Line[arrow,arrowpos=0.5,arrowlength=5,arrowwidth=2,arrowinset=0.2](288,120)(240,184)
    \Vertex(192,-24){5.657}
    \Gluon(320,24)(400,24){7.5}{6}
    \Gluon(320,-72)(400,-72){7.5}{6}
    \Line[arrow,arrowpos=0.5,arrowlength=5,arrowwidth=2,arrowinset=0.2](400,-72)(400,24)
    \Line[arrow,arrowpos=0.5,arrowlength=5,arrowwidth=2,arrowinset=0.2](400,24)(480,24)
    \Line[arrow,arrowpos=0.5,arrowlength=5,arrowwidth=2,arrowinset=0.2](480,-72)(400,-72)
    \Gluon(544,-72)(624,-72){7.5}{6}
    \Gluon(544,24)(624,24){7.5}{6}
    \Line[arrow,arrowpos=0.5,arrowlength=5,arrowwidth=2,arrowinset=0.2](624,232)(624,136)
    \Line[arrow,arrowpos=0.5,arrowlength=5,arrowwidth=2,arrowinset=0.2](624,24)(624,-72)
    \Line[arrow,arrowpos=0.5,arrowlength=5,arrowwidth=2,arrowinset=0.2](624,-72)(704,-72)
    \Line[arrow,arrowpos=0.5,arrowlength=5,arrowwidth=2,arrowinset=0.2](704,24)(624,24)
    \Vertex(400,-72){5.657}
    \Vertex(624,24){5}
    \Text(190,312)[lb]{\normalsize{\Black{SM}}}
    \Text(406,312)[lb]{\normalsize{\Black{SM}}}
    \Text(630,312)[lb]{\normalsize{\Black{SM}}}
    \Text(304,184)[lb]{\normalsize{\Black{$+$}}}
    \Text(288,-24)[lb]{\normalsize{\Black{$+$}}}
    \Text(512,-24)[lb]{\normalsize{\Black{$+$}}}
    \Text(98,488)[lt]{\normalsize{\Black{$g$}}}
    \Text(98 ,342)[lt]{\normalsize{\Black{$g$}}}
    \Text(298,488)[lt]{\normalsize{\Black{$t$}}}
    \Text(298,344)[lt]{\normalsize{\Black{$\bar t$}}}
    \Text(512,184)[lb]{\normalsize{\Black{$+$}}}
  \end{picture}
}
\end{center}
\caption{\small Feynman diagrams for $gg\to t\bar{t}$ up to $\mathscr{O}\left(\Lambda^{-2}\right)$. The dark blobs denote interactions generated by the operator $\mathcal{O}_{hg}$.}\label{Fig:gg}
\end{figure}
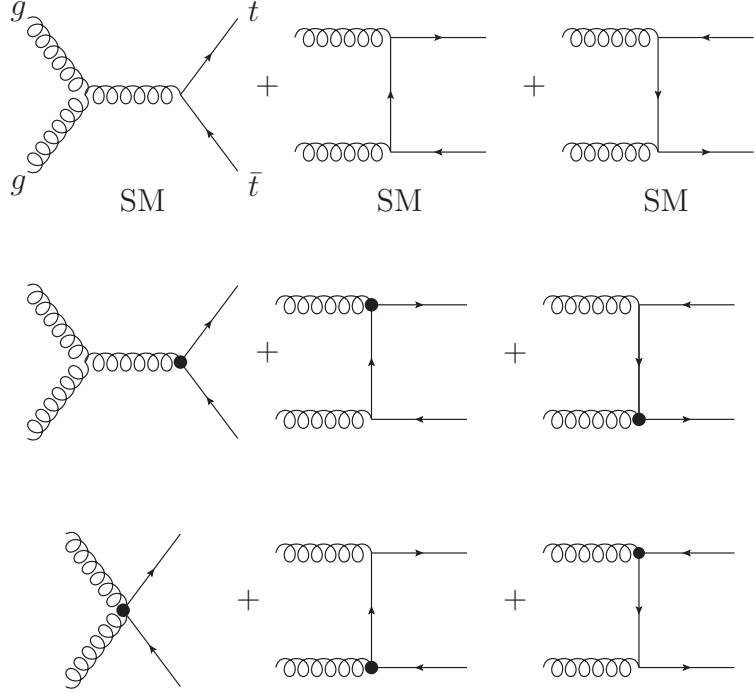
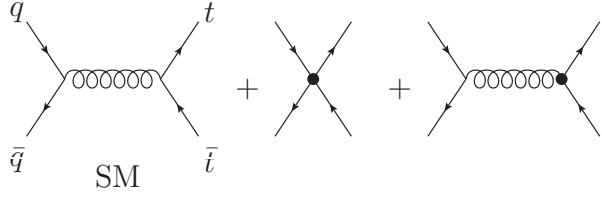
\begin{figure}[!h]
\begin{center}
\fcolorbox{white}{white}{
  \begin{picture}(528,144) (99,-91)
    \SetWidth{1.0}
    \SetColor{Black}
    \Line[arrow,arrowpos=0.5,arrowlength=5,arrowwidth=2,arrowinset=0.2](112,16)(144,-32)
    \Line[arrow,arrowpos=0.5,arrowlength=5,arrowwidth=2,arrowinset=0.2](144,-32)(112,-80)
    \Gluon(144,-32)(224,-32){7.5}{6}
    \Line[arrow,arrowpos=0.5,arrowlength=5,arrowwidth=2,arrowinset=0.2](256,-80)(224,-32)
    \Line[arrow,arrowpos=0.5,arrowlength=5,arrowwidth=2,arrowinset=0.2](224,-32)(256,16)
    \Line[arrow,arrowpos=0.5,arrowlength=5,arrowwidth=2,arrowinset=0.2](320,16)(352,-32)
    \Line[arrow,arrowpos=0.5,arrowlength=5,arrowwidth=2,arrowinset=0.2](352,-32)(320,-80)
    \Line[arrow,arrowpos=0.5,arrowlength=5,arrowwidth=2,arrowinset=0.2](352,-32)(384,16)
    \Line[arrow,arrowpos=0.5,arrowlength=5,arrowwidth=2,arrowinset=0.2](384,-80)(352,-32)
    \Line[arrow,arrowpos=0.5,arrowlength=5,arrowwidth=2,arrowinset=0.2](448,16)(480,-32)
    \Line[arrow,arrowpos=0.5,arrowlength=5,arrowwidth=2,arrowinset=0.2](480,-32)(448,-80)
    \Line[arrow,arrowpos=0.5,arrowlength=5,arrowwidth=2,arrowinset=0.2](560,-32)(592,16)
    \Line[arrow,arrowpos=0.5,arrowlength=5,arrowwidth=2,arrowinset=0.2](592,-80)(560,-32)
    \Gluon(480,-32)(560,-32){7.5}{6}
    \Text(288,-32)[lt]{\normalsize{\Black{$+$}}}
    \Text(416,-32)[lt]{\normalsize{\Black{$+$}}}
    \Text(98,32)[lt]{\normalsize{\Black{$q$}}}
    \Text(98,-90)[lt]{\normalsize{\Black{$\bar q$}}}
    \Text(262,32)[lt]{\normalsize{\Black{$t$}}}
    \Text(262,-90)[lt]{\normalsize{\Black{$\bar t$}}}
    \Text(170,-105)[lt]{\normalsize{\Black{SM}}}
    \Vertex(352,-32){5}
    \Vertex(560,-32){5}
  \end{picture}
}
\end{center}
\caption{\small Feynman diagrams for $q\bar{q}\to t\bar{t}$ up to $\mathscr{O}\left(\Lambda^{-2}\right)$. The diagram in the middle originates from the four-fermion interactions induced by the operators $\mathcal{O}_{L/R v}$, $\mathcal{O}_{L/R a}$ and $\mathcal{O}_{Qq}^{(8,3)}$. The diagram on the right is the contribution from the operator $\mathcal{O}_{hg}$.}\label{Fig:qq}
\end{figure}
%
\section{Helicity cross sections and $m_{t\bar{t}}$ distribution}
\label{app:xsec}

As explained in Section~\ref{sec:relop}, when summed over the helicities of the final top, the cross section for the $t\bar{t}$ production depends only on the sum $c_{Vv}=c_{Rv}+c_{Lv}$ (and on the suppressed isospin-1 sector contribution  $c^{\prime}_{Vv}$ defined in Eq.(\ref{eq:isospin1})). However the individual helicity cross~sections are sensitive to $c_{Rv}$ and $c_{Lv}$ individually since at high energy ${\mathcal O}_{Rv}$ (${\mathcal O}_{Lv}$) should produce mainly right (left) handed top and left (right) handed antitop.
Explicitly, the helicity cross sections are given by (we recall that $c_{Av}=c_{Rv}-c_{Lv}$)
\begin{eqnarray}
\sigma_{RR}(gg\rightarrow t\bar{t}) 
&=& \frac{\pi  \alpha_s ^2 }{24 \left(4 m^2-s\right)
   s^3}\Bigg\{\left(16 m_t^4+58 s m_t^2+s^2\right) \log \left(\frac{s+\sqrt{s \left(s-4 m_t^2\right)}}{s-\sqrt{s \left(s-4 m_t^2\right)}}\right) m_t^2 \nonumber\\
 &&\qquad\qquad\qquad\qquad-2 \sqrt{s \left(s-4 m_t^2\right)} \left(62 m_t^4-7 s m_t^2+2 s^2\right)\nonumber\\
 &&\qquad\qquad\qquad\qquad- \frac{c_{hg}}{g_s\Lambda^2} 2 \sqrt{2}  s v m_t \Bigg[\sqrt{s \left(s-4 m_t^2\right)} \left(14 m_t^2+13 s\right)\nonumber\\
 &&\qquad\qquad\qquad\qquad+\left(4 m_t^4-34 m_t^2 s\right) \log \left(\frac{s+\sqrt{s \left(s-4 m_t^2\right)}}{s-\sqrt{s \left(s-4 m_t^2\right)}}\right)\Bigg] 
   \Bigg\},\nonumber\\
\sigma_{LL}(gg\rightarrow t\bar{t}) 
&=& \sigma_{RR}(gg\rightarrow t\bar{t}),\nonumber\\
\sigma_{RL}(gg\rightarrow t\bar{t}) 
&=& \left(1+\frac{c_{hg}}{g_s\Lambda^2}4 \sqrt{2} m_t v \right)\pi  \alpha_s^2\times \nonumber\\
&&\frac{ 11 \sqrt{s \left(s-4 m_t^2\right)} \left(m_t^2-s\right)+\left(2 m_t^4-s m_t^2-4
   s^2\right) \log \left(\frac{s-\sqrt{s \left(s-4 m_t^2\right)}}{s+\sqrt{s \left(s-4 m_t^2\right)}}\right) }{24 \left(s-4 m_t^2\right) s^2},\nonumber\\
\sigma_{LR}(gg\rightarrow t\bar{t}) &=& \sigma_{RL}(gg\rightarrow t\bar{t}).   
\end{eqnarray}
and for the quark annihilation, by  (the $+$ sign in front of $c_{Vv}^{\prime}$ and ${c'_{Av}}$  is for the up quark, the $-$ for the down quark)
\begin{eqnarray}
\sigma_{RR}(q\bar{q}\rightarrow t\bar{t}) 
&=& \frac{8m_t^2\pi\alpha_s^2}{27s^{5/2}}\sqrt{s-4m_t^2}\left(1+\frac{c_{hg}}{g_s\Lambda^2}\sqrt2\frac{vs}{m_t}+
\frac{c_{Vv}\pm \frac{c_{Vv}^\prime}{2}}{g_s^2 \Lambda^2}s
\right),\nonumber\\
\sigma_{LL}(q\bar{q}\rightarrow t\bar{t}) 
&=& \sigma_{RR}(q\bar{q}\rightarrow t\bar{t}),\nonumber\\
\sigma_{RL/LR}(q\bar{q}\rightarrow t\bar{t}) 
&=& \frac{4\pi\alpha_s^2}{27s^{3/2}}\sqrt{s-4m_t^2}\bigg(1+\frac{c_{hg}}{g^2_s\Lambda^2}4\sqrt2vm_t+\frac{\sqrt{s}}{g_s^2\Lambda^2}\Big((c_{Vv}\pm \frac{c_{Vv}^\prime}{2})\sqrt{s}\nonumber\\
&&\qquad\qquad\qquad\qquad\qquad\qquad\qquad \pm (c_{Av} \pm \frac{c'_{Av}}{2})
\sqrt{s-4m_t^2}\Big)\bigg)
\end{eqnarray}
The first/second index indicates the helicity of the top/antitop. There are no effects of the operators ${\mathcal O}_{Ra}$ and ${\mathcal O}_{La}$ on the spin correlation because after integration over the variable $t$, their helicity cross~sections vanish.

When summing over the final helicities, we arrive at 
\begin{equation}
\sigma\left(q\bar{q}\rightarrow t\bar{t}\right) = \sigma_{SM}^{q\bar{q}}\left(1+
\frac{c_{Vv}\pm \frac{c_{Vv}^\prime}{2}}{g_s^2 \Lambda^2}s\right)+\frac{1}{\Lambda^2}\frac{\alpha_s}{9s^{3/2}} 4 g_s c_{hg}\sqrt{2}vm_t \sqrt{s-4m_t^2},
\end{equation}
\begin{equation}
\sigma\left(gg\rightarrow t\bar{t}\right) = \sigma_{SM}^{gg}-\frac{ vm_t\alpha_s g_s}{12\sqrt{2}\Lambda^2s^2}  c_{hg} \left(8s\log\left(\frac{s-\sqrt{s(s-4m_t^2)}}{s+\sqrt{s(s-4m_t^2)}}\right)+9\sqrt{s(s-4m_t^2)}\right),
\end{equation}
where
\begin{equation}
\sigma_{SM}^{q\bar{q}} = \frac{8 \pi \alpha_s ^2 \sqrt{s-4 m^2} \left(2 m^2+s\right) }{27 s^{5/2}},
\end{equation}
\begin{equation}
\sigma_{SM}^{gg} = \frac{\pi  \alpha_s ^2}{12 s^3} \left[{4 \left(m^4+4 s m^2+s^2\right) \log \left(\frac{s+\sqrt{s \left(s-4 m^2\right)}}{s-\sqrt{s \left(s-4 m^2\right)}}\right)-\sqrt{s \left(s-4 m^2\right)} \left(31 m^2+7 s\right)}\right].
\end{equation}
These expressions correspond to  the differential cross~sections~(\ref{qq}) and~(\ref{gg}) integrated over the scattering angle.


\providecommand{\href}[2]{#2}

\end{document}